\documentclass[twocolumn,aps,floatfix,amsmath,amssymb,showpacs,nofootinbib]{revtex4} 
\usepackage{graphicx}

\newcommand{\be}{\begin{equation}} \newcommand{\ee}{\end{equation}}
\newcommand{\bea}{\begin{eqnarray}} \newcommand{\eea}{\end{eqnarray}}
\newcommand{\bi}{\begin{itemize}} \newcommand{\ei}{\end{itemize}}
\newcommand{\bn}{\begin{enumerate}} \newcommand{\en}{\end{enumerate}}

\newcommand{\etal}{\emph{et~al.\ }}

\newcommand{\ket}[1]{\ensuremath{|{#1\rangle}}} 
\newcommand{\bra}[1]{\ensuremath{{\langle #1}|}}
\newcommand{\braket}[2]{\ensuremath{{\langle #1}|{#2 \rangle}}}
\newcommand{\ketbra}[2]{\ensuremath{|{#1 \rangle}{\langle #2}|}}
\newcommand{\e}{\ensuremath{\mathrm{e}}}

\begin{document}

\title{Experimental motivation and empirical consistency \\in minimal
  no-collapse quantum mechanics}

\author{Maximilian Schlosshauer}

\email{MAXL@u.washington.edu}

\affiliation{Department of Physics, University of Washington, Seattle,
  Washington 98195, USA}


\begin{abstract}
  
  We analyze three important experimental domains (SQUIDs, molecular
  interferometry, and Bose-Einstein condensation) as well as
  quantum-biophysical studies of the neuronal apparatus to argue that
  (i) the universal validity of unitary dynamics and the superposition
  principle has been confirmed far into the mesoscopic and macroscopic
  realm in all experiments conducted thus far; (ii) all observed
  ``restrictions'' can be correctly and completely accounted for by
  taking into account environmental decoherence effects; (iii) no
  positive experimental evidence exists for physical state-vector
  collapse; (iv) the perception of single ``outcomes'' is likely to be
  explainable through decoherence effects in the neuronal apparatus.
  We also discuss recent progress in the understanding of the
  emergence of quantum probabilities and the objectification of
  observables. We conclude that it is not only viable, but moreover
  compelling to regard a minimal no-collapse quantum theory as a
  leading candidate for a physically motivated and empirically
  consistent interpretation of quantum mechanics.

\end{abstract}

\maketitle

\section{Introduction}

Historically, quantum theory was motivated by the need to describe the
behavior of microscopic systems not explainable by the laws of
classical physics. Not only was quantum mechanics deemed unnecessary
for a description of the macroworld of our experience, it also led to
``strange'' consequences that seemed to blatantly contradict our
experience, as famously illustrated by the Schr\"odinger-cat {\em
  Gedanken} experiment \cite{Schrodinger:1935:gs} and later generally
referred to as the ``measurement problem.'' Therefore quantum theory
was often, as in the Copenhagen interpretation, banned {\em a priori}
from the macrosopic realm.

Over the past decade, however, a rapidly growing number of experiments
have demonstrated the existence of quantum superpositions of
mesoscopically and macrosopically distinct states on increasingly
large scales. Such superpositions are observed as individual quantum
states and are perfectly explained by unitarily evolving wave
functions.  On the other hand, decoherence theory
\cite{Zeh:1970:yt,Zeh:1973:wq,Zurek:1981:dd,Zurek:2002:ii,Joos:2003:jh,Schlosshauer:2003:tv}
has enabled one to understand the fragility of such superpositions,
and thus the extreme difficulty in observing them outside of
sophisticated experimental setups, as being due to ubiquitous quantum
interactions with environmental degrees of freedom.

These developments have thus extended the domain for an application of
quantum theory far into the mesoscopic and macroscopic realm, which
lends strong support to assuming a universally exact and applicable
Schr\"odinger equation. To make a physically compelling case for such
a purely unitary quantum theory we must pursue two related goals.
First, we ought to continue to design experiments which demonstrate
the existence of quantum superpositions of macrosopically distinct
states --- and which, ideally, can explicitly rule out collapse
models. Second, since the assumption of a universal Schr\"odinger
dynamics implies that superpositions of (presumably macroscopically)
different observer states are both possible and inescapable if we
include physical observers into the quantum-mechanical description, we
must simultaneously show that environmental decoherence provides the
necessary and sufficient mechanism to explain our observation of a
``classical'' world. The emergence of the latter can then be
understood not only in spite of, but precisely {\em because} of the
quantum formalism --- no classical prejudice need to be imposed.

The formal basis for a derivation of a viable interpretation of
quantum mechanics from the ``bare'' unitary formalism alone has been
outlined in several papers. The basic idea was introduced in Everett's
proposal of a relative-state view of quantum mechanics
\cite{Everett:1957:rw}. It was later adapted and popularized by deWitt
\cite{DeWitt:1970:pl,DeWitt:1971:pz,DeWitt:1973:pz} in his
``many-worlds'' interpretation of quantum mechanics, whose elements go
far beyond the abstract sketches of Everett und which must therefore
be strictly distinguished from Everett's proposal \cite{Kent:1990:nm}.
Relative-state interpretations were subsequently fleshed out, by
taking into account decoherence effects, in works by Zeh
\cite{Zeh:1970:yt,Zeh:1973:wq,Zeh:2000:rr}, Zurek
\cite{Zurek:1998:re,Zurek:2002:ii,Zurek:2004:yb}, Wallace
\cite{Wallace:2003:iz,Wallace:2003:iq}, and others (see, for example,
\cite{Deutsch:1985:rx,Vaidmain:1998:zp,Donald:1999:yb}). Such a theory
can be based on the most minimal set of assumptions about the quantum
formalism and its interpretation.  First, a completely known (pure)
state of an isolated quantum system $\mathcal{S}$ is described by a
normalized state vector $\ket{\psi}$ in a Hilbert space
$H_\mathcal{S}$. Second.  the time evolution of a state vector
$\ket{\psi}$ is given by the Schr\"odinger equation $\mathrm{i}\hbar
\frac{\partial}{\partial t} \ket{\psi} = \widehat{H}_\mathcal{S}$,
where $\widehat{H}_\mathcal{S}$ is the Hamiltonian of the system
$\mathcal{S}$. No mention is made of measurements in this formulation.
Instead, measurements are described without special axioms in terms of
physical interactions between systems described by state vectors (wave
functions) and governed by suitable interaction Hamiltonians.
Observables then emerge as a derived concept
(see, for example, \cite{Joos:2003:jh,Zurek:2002:ii}).

In this paper, however, we take a less formal route and focus on an
analysis of the experimental and theoretical progress (with an
emphasis on the former) towards the two goals mentioned before,
namely, the continued acquisition of experimental evidence for
superpositions of macrosopically distinct states and an explanation
for the emergence of definite perceptions in spite of an assumed
universal validity of the superposition principle.

Our goal is to show that there is no experimental evidence for a
breakdown of the superposition principle and the related interference
effects at any length scale investigated thus far.  Whenever a decay
of such superpositions is observed, it can be fully accounted for
(both experimentally and theoretically) as resulting from
environmental interactions. The absense of any empirical evidence for
nonlinear deviations from unitary time evolution, combined with the
ability to give an empirically adequate description of the decoherence
of superpositions into apparent mixtures, provides good reasons to
take the universal validity of the Schr\"odinger equation as a working
assumption and to explore the consequences of this assumption.

The resulting theory will require more attention to a detailed
quantum-mechanical description of observers and observations. Such an
account is interpretation-neutral, while the question of its relevance
for solving the measurement problem may depend on the particular
features of an interpretation. This is so because there exist
interpretations, for example, Bohmian mechanics or modal
interpretations, that claim to solve the measurement problem
\emph{without} having to give an explicit account of the physical
processes describing observers and observations (see also
Sec.~\ref{sec:physcoll}).

This paper is organized as follows. In Sec.~\ref{sec:experiments}, we
shall discuss and analyze three important experimental
domains---superconducting quantum interference devices (SQUIDs),
matter-wave interferometry, and Bose-Einstein condensation---that have
provided evidence for superpositions of macroscopically
distinguishable states. Sec.~\ref{sec:physcoll} comments on the
current status of physical collapse theories in view of the described
experiments.  In Secs.~\ref{sec:prob} and \ref{sec:object}, we shall
discuss steps towards the resolution of two issues that have often
been considered as posing a challenge to relative-state
interpretations: The question of the origin of quantum probabilities
and the connection with Born's rule, and the problem of the
``objectification'' of observables and thus the emergence of
``classical reality.''  Sec.~\ref{sec:brain} analyzes theoretical
models for decoherence in the perceptive and cognitive apparatus, and
the implications of such decoherence processes.  Finally, in
Sec.~\ref{sec:discussion}, we shall summarize our main conclusions and
discuss possible next steps.

\section{Superpositions of macroscopically
  distinct states: Experiments and
  implications}\label{sec:experiments}

In the following, we shall describe three recent experimental areas
that have led to (or that are very close to achieving) the observation
of superpositions of mesoscopically and macroscopically
distinguishable states: Coherent quantum tunneling in SQUIDs
(Sec.~\ref{sec:squid}), diffraction of C$_{70}$ (and larger) molecules
in matter-wave interferometers (Sec.~\ref{sec:mol-interference}), and
number-difference superpositions in two-species Bose-Einstein
condensates (Sec.~\ref{sec:bose}).  These experiments have achieved
the largest such superpositions observed thus far and also represent
the most promising experimental domains for achieving even larger
superpositions in the future.

For some earlier experiments demonstrating mesoscopic and macrosopic
quantum effects, see the setups using superconductors
\cite{Clarke:1988:tv,Rouse:1995:yb,Silvestrini:1997:ra,Rouse:1998:om,Nakamura:1999:ub},
nanoscale magnets
\cite{Friedman:1996:om,Wernsdorfer:1997:qq,Barco:1999:tb},
laser-cooled trapped ions \cite{Monroe:1996:tv}, and photons in a
microwave cavity \cite{Brune:1996:om,Raimond:1997:um}. We would also
like to mention Leggett's review article \cite{Leggett:2002:uy} which
discusses some experiments that probe the limits of quantum mechanics.
Leggett's motivation, however, is somewhat different than that of the
present author, as Leggett's main aim is to assess the status of physical
collapse theories in view of these experiments.

\subsection{Measuring the macrosopic distinctness of states
  in a superposition} \label{sec:measures}

Before embarking on an analysis of the experiments, we shall first
lend a more precise meaning to the ubiquitous phrase ``superposition
of macrosopically distinct (or distinguishable) states.'' If
confronted with a superposition of two states $\ket{A}$ and $\ket{B}$
of the form
\be
\ket{\Psi} = \frac{1}{\sqrt{2}} \bigl( \ket{A} + \ket{B} \bigr),
\ee
how are we to decide whether this indeed represents a macrosopic
Schr\"odinger-cat state? Clearly, two conditions will need to be
fulfilled:

\bn

\item The states $\ket{A}$ and $\ket{B}$ must differ macrosopically in
  some extensive quantity (e.g., spatial separation, total mass,
  magnetic moment, momentum, charge, current, etc.), relative to a
  suitable microsopic reference value.
  
\item The degree of GHZ-type entanglement \cite{Greenberger:1990:bw}
  in the state $\ket{\Psi}$, i.e., the number of correlations that
  would need to be measured in order to distinguish this state from a
  mixture, must be sufficiently large.  With $\ket{A}$ and $\ket{B}$
  usually representing GHZ-like multi-particle states in complex
  systems such as superconducting currents, molecules, and atomic
  gases, this measure can typically be well-estimated by the number of
  microsopic constituents (electrons, protons, neutrons) in the
  system.

\en

A similiar combination of two measures has been suggested by Leggett
\cite{Leggett:1980:yt,Leggett:2002:uy} under the labels ``extensive
difference'' and ``disconnectivity.'' We shall adopt Leggett's former
term for the first condition, and use the term ``degree of
entanglement'' for the second. A both necessary and sufficient
condition for a superposition to be considered a superposition of
macroscopically distinct states is then given by the requirement that
both the extensive difference between $\ket{A}$ and $\ket{B}$ and the
interparticle entanglement in $\ket{\Psi}$ be large relative to an
appropriate microsopic unit.

\subsection{Superconducting quantum interference
  devices} \label{sec:squid}

Experiments using SQUIDs have not only demonstrated that the dynamics
of a macrosopic quantity of matter (here $\approx 10^9$ Cooper pairs)
can be collectively determined by a single macrosopic coordinate
governed by quantum mechanics, but have also achieved the creation and
indirect observation of quantum superpositions of two truly macrosopic
states that correspond to currents of several $\mu$A running in
opposite directions.

\subsubsection{SQUID setup and detection of superpositions of
  macroscopically distinct currents}

A SQUID consists of a superconducting loop interrupted by a Josephson
junction and immersed into an external magnetic field that creates a
flux $\Phi_\text{ext}$ through the loop. This allows for a persistent
dissipationless current (``supercurrent'') to flow around the loop, in
clockwise or counterclockwise direction, creating an additional flux.
Such a current is composed of a very large number of Cooper pairs
(i.e., Bose-condensed electron pairs) whose collective center-of-mass
motion can be described by a macrosopic wave function around the loop.

Since the wave function must be continuous around the loop, an integer
$k$ times its wavelength must equal the circumference of the loop.
Since the Josephson junction induces a discontinuous phase drop
$\Delta \phi_J$, and since the total change in phase around the
superconducting loop is given by $2\pi\Phi/\Phi_0$, where
$\Phi_0=h/2e$ is the flux quantum and $\Phi$ is the total trapped flux
through the loop, the phase continuity condition implies
\be
\Delta \phi_J + 2\pi \Phi/\Phi_0 = 2\pi k,
\ee
with $k=1,2,\cdots$. This means that the collective quantum dynamics
of the SQUID are determined by the single macrosopic variable
$\Phi$.

The effective SQUID Hamiltonian can be written as \cite{Weiss:1999:tv}
\begin{multline} \label{eq:squid-h}
\widehat{H} = \frac{\widehat{P}_\Phi^2}{2C} + U(\Phi) 
= - \frac{\hbar^2}{2C}
 \frac{d^2}{d\Phi^2} + \biggl[ \frac{(\Phi- 
  \Phi_\text{ext})^2}{2L} \\ - \frac{I_c \Phi_0}{2\pi} \cos \biggl(
2\pi \frac{\Phi}{\Phi_0} \biggr) \biggr],
\end{multline}
where $C$ is the total capacitance (mainly due to the junction), $L$
is the (finite) self-inductance of the loop, and $I_c$ is the critical
current of the junction. This Hamiltonian induces dynamics that are
analogous to the motion of a particle with effective ``mass'' $C$
moving in $\Phi$-space in a tilted one-dimensional double-well
potential, with the tilt determined by $\Phi_\text{ext}$. The role of
the canonical variables $\widehat{X}$ and $\widehat{P}$ is here played
by the total trapped flux $\widehat{\Phi}$ and the total displacement
current $\widehat{P}_\Phi = -\mathrm{i}\hbar d/d\widehat{\Phi}$ (which has
units of charge; $Cd\widehat{P}_\Phi / dt$ is the charge difference
across the junction).

A set of eigenstates $\ket{k}$ of the Hamiltonian of
Eq.~\eqref{eq:squid-h}, called ``$k$-fluxoid states,'' are localized
in one of the wells of the potential below the (classically
impenetrable) barrier if the damping induced by the Josephson junction
is weak. The corresponding wave functions $\psi_k(\Phi) \equiv
\braket{\Phi}{k}$ are locally $s$-harmonic, so their amplitudes are
peaked around the respective minimum of $U(\Phi)$ with narrow spreads
in flux space. Thus these low-lying energy eigenstates have a
relatively small range of associated flux values and can therefore (at
least for sufficiently small $k$) also be viewed as ``fuzzy''
eigenstates of the flux operator. By adjusting $\Phi_\text{ext}$, the
energy levels are shifted, and for certain values of
$\Phi_\text{ext}$, two levels in opposite wells can be made to align,
which allows for resonant quantum tunneling between the wells (i.e.,
between two fluxoid states) \cite{Silvestrini:1996:ii,Rouse:1998:om},
leading to a macroscopic change in the magnetic moment of the system.

The most important states for our subsequent treatment are the
zero-fluxoid state $\ket{0}$ and the one-fluxoid state $\ket{1}$.
Since the states $\ket{0}$ and $\ket{1}$ are localized in,
respectively, the left and right well of the potential, let us denote
them by $\ket{L}$ and $\ket{R}$ in the following. These states
correspond (apart from the quantum zero-point energy
\cite{Wal:2000:om}) to a classical persistent-current state and thus
to macrosopically distinguishable directions of the superconducting
current.  Since other states are well-separated in energy, the SQUID
can thus be effectively modelled as a macroscopic quantum-mechanical
two-state system (i.e., as a macrosopic qubit).

\begin{figure}
\begin{center}
\includegraphics[scale=0.4]{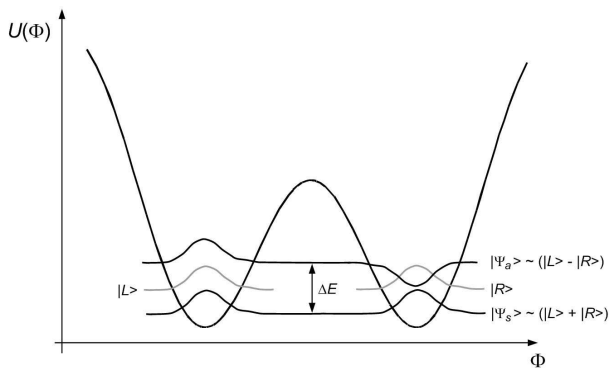}
\end{center}
\caption[Effective SQUID potential at bias $\Phi_\text{ext} = \Phi_0/2$]{\label{fig:squid-pot}
  Effective SQUID potential at bias $\Phi_\text{ext} = \Phi_0/2$. At
  this point, the double-well potential becomes symmetric. The
  degeneracy between the two fluxoid states $\ket{L}$ and $\ket{R}$
  (which are localized in the left and right well of the potential and
  correspond to macrosopic currents running in opposite direction
  around the loop) is lifted by the formation of delocalized coherent
  superpositions $\ket{\Psi_s} = \frac{1}{\sqrt{2}} \bigl( \ket{L} +
  \ket{R} \bigr)$ (the symmetric ground state) and $\ket{\Psi_a} =
  \frac{1}{\sqrt{2}} \bigl( \ket{L} - \ket{R} \bigr)$ (the
  antisymmetric first excited state).  The energy difference $\Delta
  E$ between $\ket{\Psi_s}$ and $\ket{\Psi_a}$ has been experimentally
  measured \cite{Friedman:2000:rr,Wal:2000:om}, which confirms the
  existence of superpositions of the macroscopically distinct states
  $\ket{L}$ and $\ket{R}$.}
\end{figure}

At bias $\Phi_\text{ext} = \Phi_0/2$, the well becomes symmetric and
the corresponding two fluxoid states $\ket{L}$ and $\ket{R}$ would
become degenerate (see Fig.~\ref{fig:squid-pot}).  However, the
degeneracy is lifted by the formation of symmetric and antisymmetric
superpositions of $\ket{L}$ and $\ket{R}$ that represent the new
energy ground state,
\be
\ket{\Psi_s} = \frac{1}{\sqrt{2}} \bigl( \ket{L} + \ket{R} \bigr)
\ee
with energy $E_+$, and the first excited energy eigenstate
\be
\ket{\Psi_a} = \frac{1}{\sqrt{2}} \bigl( \ket{L} - \ket{R} \bigr)
\ee
with energy $E_-$.  Thus these eigenstates are delocalized across the
two wells. The (typically very small) energy splitting $\Delta E = E_a
- E_s$ is determined by the WKB matrix elements for tunneling between
the two wells (and thus between $\ket{L}$ and $\ket{R}$), and is only
dependent on the capacitance $C$ of the junction, scaling as $\Delta E
\propto \e^{-\sqrt{C}}$.    

\begin{figure}
\begin{center}
\includegraphics[scale=.25]{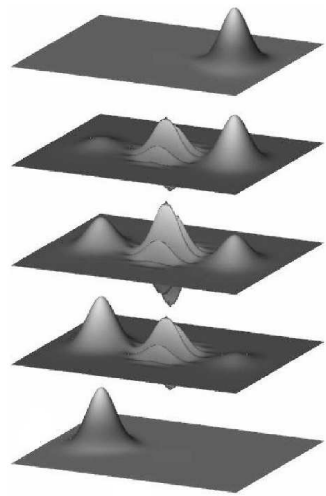}
\end{center}
\caption[Time evolution of the Wigner function
corresponding to a superposition $\ket{\Psi(t)} \propto \ket{L} \cos
(\Delta E t / 2) + i \ket{R} \sin(\Delta E t / 2)$ of the two
localized opposite-current states $\ket{L}$ and $\ket{R}$ in a
SQUID]{\label{fig:wigner1} Time evolution of the Wigner function
  corresponding to a superposition $\ket{\Psi(t)} \propto \ket{L} \cos
  (\Delta E t / 2) + i \ket{R} \sin(\Delta E t / 2)$ of the two
  localized opposite-current states $\ket{L}$ and $\ket{R}$ in a
  SQUID.  The state coherently oscillates between the two wells,
  leading to coherent quantum tunneling. This manifests itself in a
  macrosopic current oscillating between clockwise and
  counterclockwise directions.  Figure reprinted with permission from
  \cite{Everitt:2004:mb}. Copyright 2004 by the American Physical
  Society.  }
\end{figure}

If the system is now more generally described by an arbitrary
superposition of $\ket{L}$ and $\ket{R}$, $\ket{\Psi(t)}= \alpha(t)
\ket{L} + \beta(t) \ket{R}$, and if we choose the left-localized state
$\ket{L}$ as the initial state of the SQUID, i.e., $\ket{\Psi(t=0)} =
\ket{L}$, we obtain the time evolution
\be
\ket{\Psi(t)} \propto \ket{L} \cos (\Delta E t / 2) + i \ket{R}
  \sin(\Delta E t / 2). 
\ee
Thus the wave function oscillates coherently between the two localized
current states $\ket{L}$ and $\ket{R}$ in each well (see
Fig.~\ref{fig:wigner1}) at a rate determined by $\Delta E$, since the
probability to find the wave function localized in, say, the left well
is oscillatory in time,
\be
P_L(t) = | \braket{L}{\Psi(t)} |^2 = \cos^2 (\Delta E t / 2).
\ee
This leads to coherent quantum tunneling between the two wells and
manifests itself in an oscillation of the current in the SQUID between
clockwise and counterclockwise directions. This tunneling effect has
been directly observed in superconducting qubit setups similiar to the
one described here
\cite{Nakamura:1999:ub,Korotkov:2001:mq,Korotkov:2001:my,Greenberg:2002:mi,%
  Martinis:2002:qq,Yu:2002:yb,Vion:2002:oo}.  

The indirect route for detecting the presence of superpositions of states
corresponding to macrosopic currents running in opposite directions
relies on a static spectroscopic measurement of the energy difference
$\Delta E$ (see Fig.~\ref{fig:squid-pot}). Friedman \etal
\cite{Friedman:2000:rr} have confirmed the existence of such an energy
gap (in excellent agreement with theoretical predictions) and,
therefore, of superpositions of macroscopically distinct fluxoid states
(see also \cite{Wal:2000:om} for a similiar experiment and
result). In their setup, $\ket{L}$ and $\ket{R}$ (which in this
experiment corresponded to $k=4$ and $k=10$, respectively) differed in
flux by more than $\Phi_0/4$ and in current by 2--3~$\mu$A,
corresponding to about $10^{10} \mu_B$ in local magnetic moment.
Furthermore, the dynamics of the in-unison motion of the approximately
$10^9$ Cooper pairs represented by $\ket{L}$ and $\ket{R}$ are given
by a single unitarily evolving wave function representing the
collective flux coordinate $\Phi$.

\subsubsection{Scaling}

A main advantage of SQUIDs over other experiments (such as those
described in the subsequent sections) that probe the limits of quantum
mechanics lies in the fact that the relevant macrosopic variable,
namely, the trapped flux through the SQUID ring, can be controlled by
means of microsopic energy differences in the Josephson junction
\cite{Leggett:2002:uy}. As mentioned before, the tunneling matrix
element scales as $\e^{-\sqrt{C}}$, where $C$ is dominantly
determined by the junction rather than by the size of the loop. Thus
the difficulty of observing superpositions of macrosopically distinct
states scales essentially independently of the degree of macrosopic
distinctness between these states (i.e., difference in flux between
the opposite currents). This is in stark contrast to the matter-wave
diffraction experiments and Bose-Einstein condensates discussed below.
In the first case, the grating spacing must decrease as $1/\sqrt{N}$
with the number $N$ of atoms in the molecule, in the second case the
decoherence rate increases as $N^2$ with the number $N$ of atoms in the
condensate.

This particular property of SQUIDs has allowed for the creation of
superpositions of states that differ by several orders of magnitude
more than in other experiments (see Sec.~\ref{sec:scaling} below).

\subsubsection{The interpretation of superpositions}  \label{sec:interpret-superpos}

It is well known that quantum-mechanical superpositions must not be
interpreted as a simple superposition (addition) of probability
distributions.  Formally, this conclusion is of course well-reflected
in the fact that, in quantum mechanics, we deal with superpositions of
probability amplitudes rather than of probabilities, leading to
interference terms in the probability distribution.

However, this crucial difference between classical and
quantum-mechanical superpositions is sometimes not sufficiently
clearly brought out when describing particular experimental
situations. In the case of the standard double-slit interference
experiment, for example, the state of the diffracted particle is
described by a coherent superposition $\ket{\psi} = \bigl(
\ket{\psi_1} + \ket{\psi_2} \bigr) / \sqrt{2}$ of the states
$\ket{\psi_1}$ and $\ket{\psi_2}$ corresponding to passage through
slit 1 and 2, respectively. This is frequently interpreted as simply
representing simultaneous passage of the particle through both slits,
i.e., presence of the particle in two distinct spatial regions at the
same time, thereby tacitly neglecting the interference terms in the
probability distribution.

In the double-slit example, this view will not necessarily be
disproven until the stage of the screen is reached at which
interference fringes appear. Similiarly, and even more drastically,
the superpositions of macrosopically distinct current states in a
SQUID show that the simplified view of a classical superposition of
probability distributions is inadequate. For, if this view were
correct, the two contributing opposite currents would mutually cancel
out and thus the net ``current'' described by this superposition would
have to be zero, contrary to what is observed.  Instead, the SQUID
opposite-current superposition represents a novel individually
existing physical state that can be described as a coherent
``interaction'' between simultaneously present states representing
currents of opposite direction.

The SQUID example also shows that the ``splitting'' often referred to
in an Everettian framework (for example, in deWitt's popularization
of the ``many-worlds view''
\cite{DeWitt:1970:pl,DeWitt:1971:pz,DeWitt:1973:pz}) should not be
taken too literally.  The transition, i.e., the ``split,'' from a
single ``classical'' state---i.e., classically defined definite
structures such as particles (defined as having a definite position),
currents (defined as a flow of charge into a definite direction),
etc.---into a state describing a superposition of such states occurs
in a completely unitary and thus reversible manner by changing
$\Phi_\text{ext}$. There is only one single global state vector
$\ket{\Psi(t)}$ at all times that corresponds to ``physical reality.''
The decomposition into a superposition of other states is a primarily
formal procedure useful in revealing the physical quantities of
our experience contained in the arbitrary state vector
$\ket{\Psi(t)}$, since the latter can in general not be related to any
``classical'' physical structure that would correspond to directly
observed objects or properties. In this sense, the ``split'' is simply
a consequence of trying to trace throughout time a particular (usually
``classical'') state that does not coincide with $\ket{\Psi(t)}$.
Quantum mechanics shows that this can, in general, only be done in a
relative-state sense.

The decomposition obtains also {\em physical} meaning when the
dynamical evolution of the system described by $\ket{\Psi(t)}$ is
considered, as the coefficients multiplying the ``classical'' terms in
the superposition will in general be time-dependent. In the example of
the SQUID, the coherent-tunneling state does not directly relate to a
current in the classical sense (i.e., a current of definite
direction), but it can be decomposed into two such currents of
opposite direction. The physical relevance of this decomposition and
the meaning of the superposition then manifests itself as a current
that oscillates between clockwise and counterclockwise directions.

\subsubsection{Decoherence and the preferred basis} \label{sec:squid-prefbasis}

A particularly interesting feature of the macrocurrent superpositions
in SQUIDs is the fact that the interaction with the environment leads
to a localization in flux space, rather than to the much more familiar
and common localization in position space. In other words, the
``preferred basis'' (Zurek's ``pointer states''
\cite{Zurek:1981:dd,Zurek:1982:tv}) of the SQUID are flux eigenstates.

This observation is perfectly well accounted for by decoherence
theory, which describes the selection of the preferred basis by means
of the stability criterion, first formulated by Zurek
\cite{Zurek:1981:dd} (see also
\cite{Zurek:1982:tv,Zurek:1993:pu,Zurek:1998:re,Zurek:2002:ii,Schlosshauer:2003:tv}).
According to this criterion, the basis used to represent the possible
states of the system must allow for the formation of dynamically
stable system-environment correlations.  A sufficient (albeit not
necessary) requirement for this criterion to be fulfilled is given by
the condition that all basis projectors $\widehat{P}_n =
\ketbra{s_n}{s_n}$ of the system must (at least approximately) commute
with the system-environment interaction Hamiltonian
$\widehat{H}_\text{int}$, i.e.,
\be \label{eq:commut}
[\widehat{H}_\text{int}, \widehat{P}_n] = 0 \qquad \text{for all $n$.}
\ee
That is, the preferred basis of the system is given by a set of
eigenvectors of $\widehat{H}_\text{int}$.
  
In the case of the SQUID experiments at bias $\Phi_\text{ext} =
\Phi_0/2$, if the interaction with the environment is very weak and thus
the dynamics of the SQUID system are dominantly governed by the
effective SQUID Hamiltonian $\widehat{H}$, Eq.~\eqref{eq:squid-h}, the
preferred states are predicted to be eigenstates of this Hamiltonian,
namely, the dislocalized coherent superpositions $\ket{\Psi_s} =
\frac{1}{\sqrt{2}} \bigl( \ket{L} + \ket{R} \bigr)$ and $\ket{\Psi_a}
= \frac{1}{\sqrt{2}} \bigl( \ket{L} - \ket{R} \bigr)$ of the localized
zero-fluxoid and one-fluxoid states $\ket{L}$ and $\ket{R}$. This is
in agreement both with the observation of coherent quantum tunneling
between the wells and with the measurement of the energy gap $\Delta E
= E_a - E_s$ between the states $\ket{\Psi_s}$ and $\ket{\Psi_a}$.

Under realistic circumstances, however, the SQUID is coupled to a
dissipative environment $\mathcal{E}$ which can quite generally be
modeled as a harmonic heat bath of bosons \cite{Weiss:1999:tv}, i.e.,
as a bath of $N$ harmonic oscillators with generalized coordinates
$x_\alpha$ and $p_\alpha$, natural frequency $\omega_\alpha$, mass
$m_\alpha$, and Hamiltonian
\be
\widehat{H}_\mathcal{E} = \frac{1}{2} \sum_{\alpha=1}^N \biggl(
\frac{p^2_\alpha}{m_\alpha} + m_\alpha \omega^2_\alpha x^2_\alpha
\biggr).
\ee
The reservoir modes $x_\alpha$ couple dynamically to the total flux
variable $\Phi$ of the SQUID ring. More precisely, they couple to the
fluxoid (and essentially opposite-current) states $\ket{L}$ and
$\ket{R}$ via the interaction Hamiltonian \cite{Weiss:1999:tv}
\be
\widehat{H}_\text{int} = - \sigma_z
\biggl( \frac{\varphi_0}{2} \sum_\alpha c_\alpha x_\alpha 
\biggr), 
\ee
where $\sigma_z = \bigl( \ketbra{L}{L} - \ketbra{R}{R} \bigr)$ is the
so-called ``pseudospin'' operator (owing its name to the fact that the
SQUID double-well system can be effectively mapped onto a two-state
spin system, with $\ket{L}$ and $\ket{R}$ corresponding to, say, spin
``up'' and ``down,'' respectively), and $\pm \varphi_0$ are the flux
values associated with the two localized states $\ket{L}$ and $\ket{R}$.
 
According to the commutativity criterion, Eq.~\eqref{eq:commut}, the
stable states into which the system decoheres are then eigenstates of
$\sigma_z$, i.e., the preferred basis of the system is given by the
two states $\ket{L}$ and $\ket{R}$, This, of course, is in full
agreement with observations and explains the localization in flux
space, i.e., the rapid reduction of the superposition into an apparent
ensemble of the macroscopically distinguishable current states
$\ket{L}$ and $\ket{R}$.

\begin{figure}
\begin{center}
 \includegraphics[scale=.2]{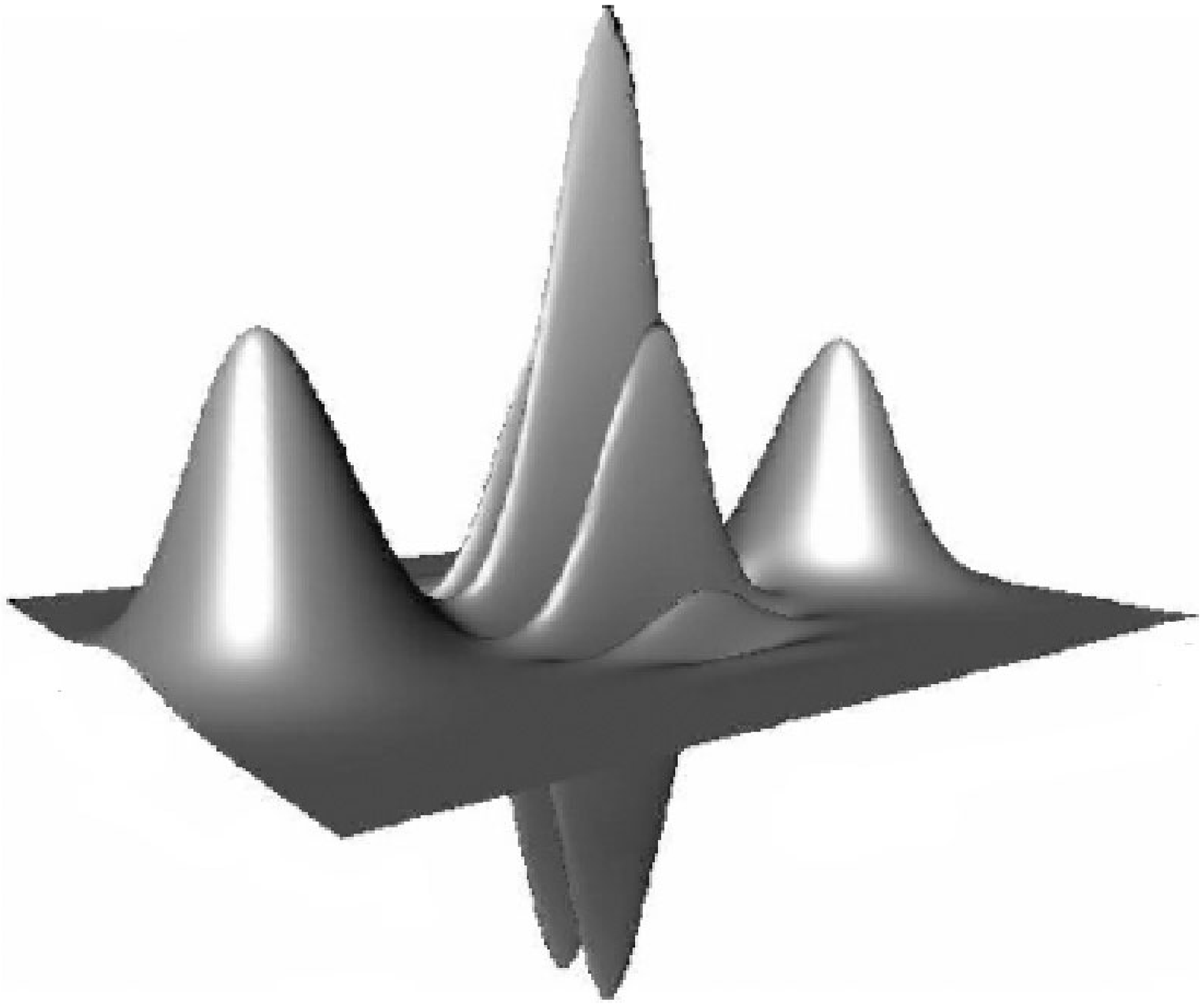}

 \includegraphics[scale=.2]{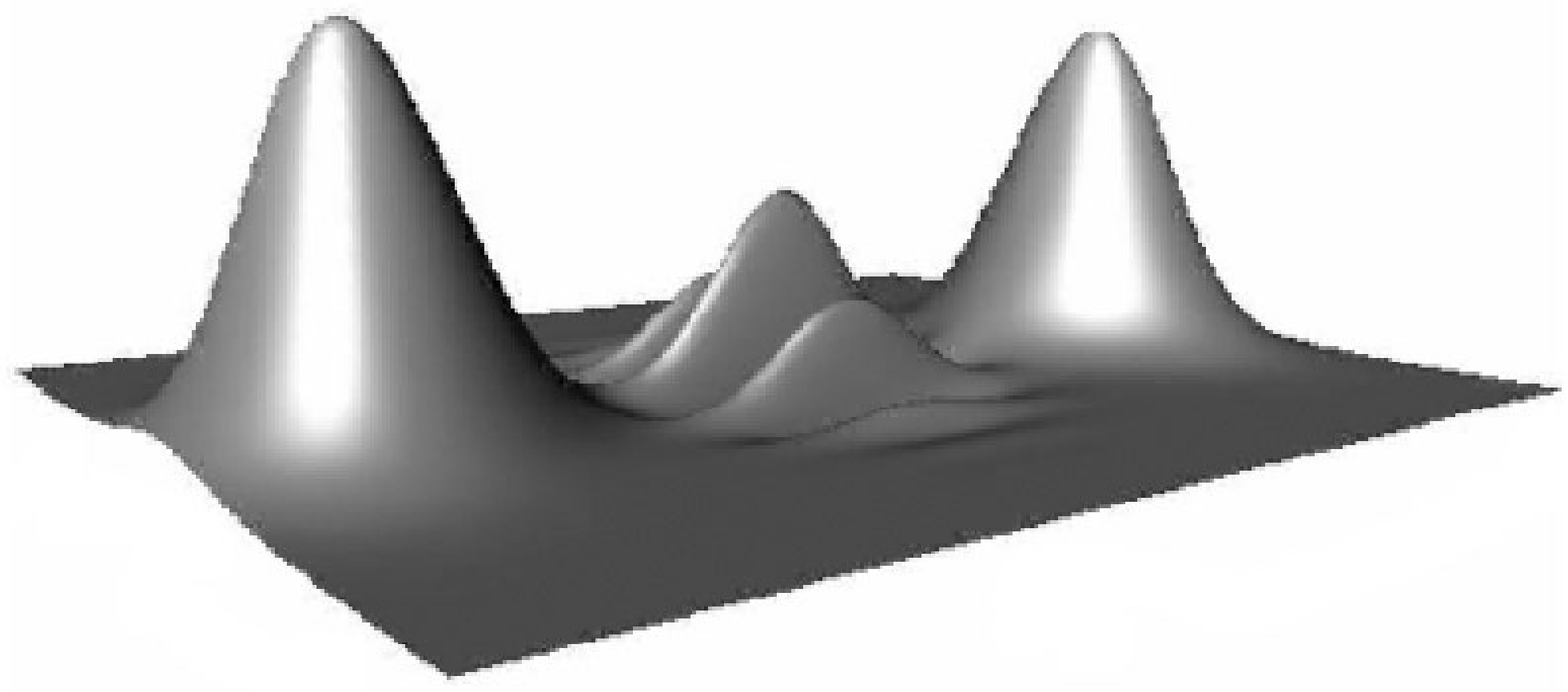}

 \includegraphics[scale=.2]{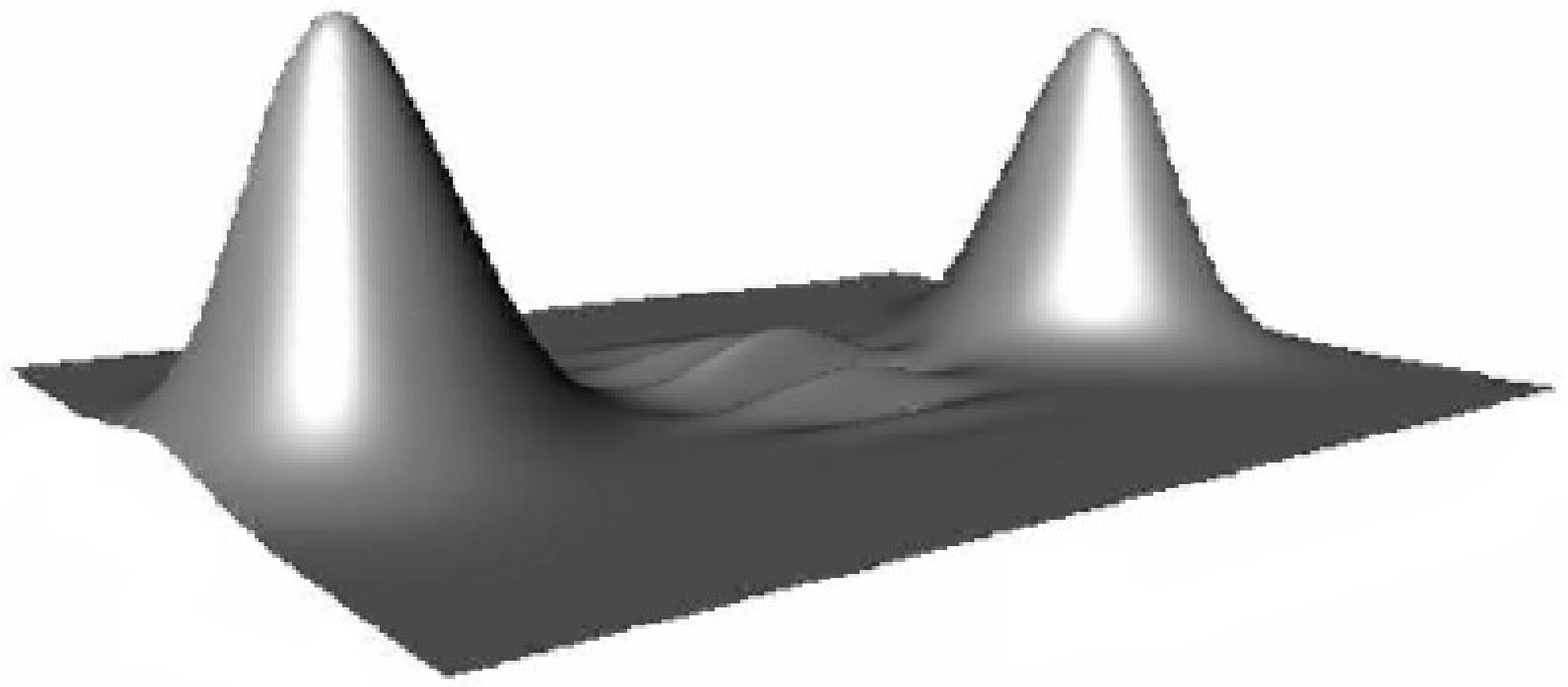}

 \includegraphics[scale=.2]{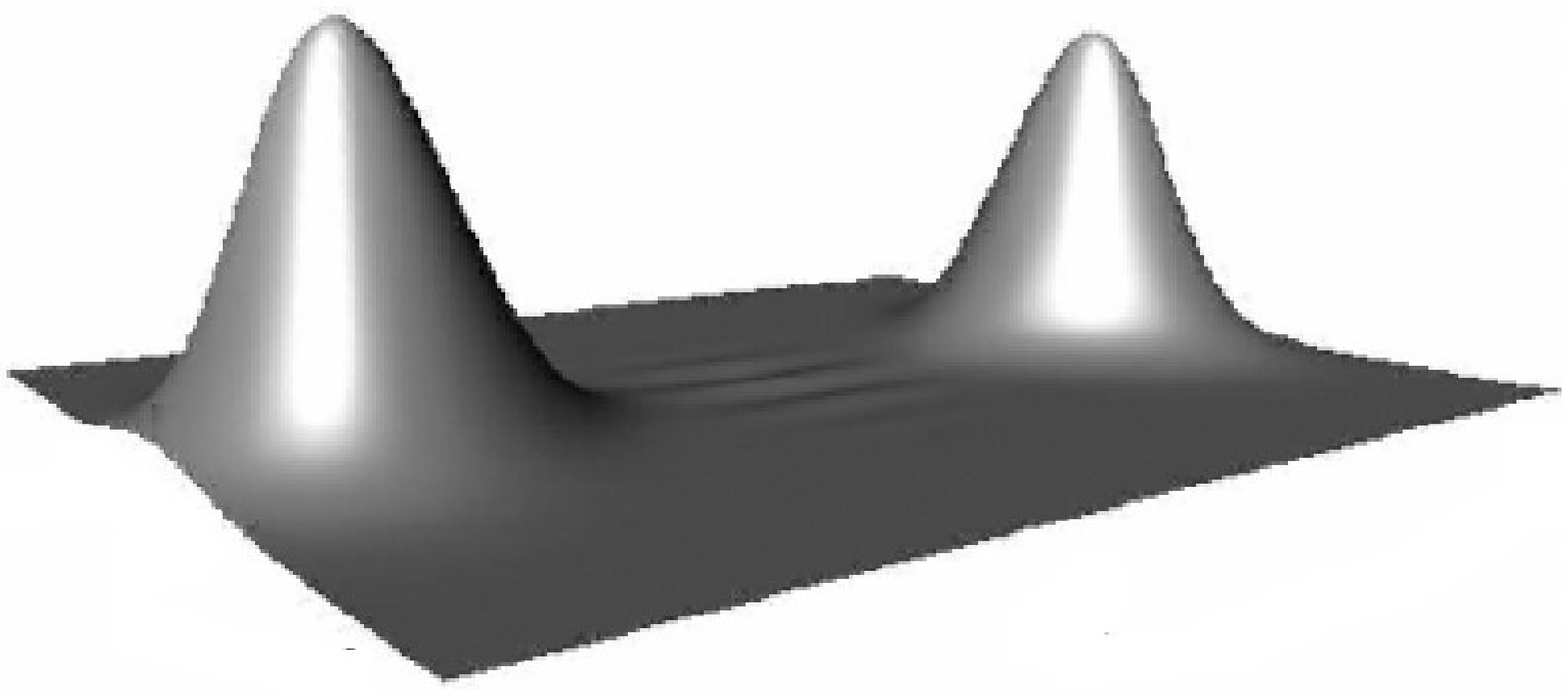}
\end{center}
\caption[Decoherence of the symmetric
ground state $\ket{\Psi_s} = \frac{1}{\sqrt{2}} \bigl( \ket{L} +
\ket{R} \bigr)$ at bias $\Phi_\text{ext} = \Phi_0/2$ in a
SQUID]{\label{fig:squid-dec}Decoherence of the symmetric ground state
  $\ket{\Psi_s} = \frac{1}{\sqrt{2}} \bigl( \ket{L} + \ket{R} \bigr)$
  at bias $\Phi_\text{ext} = \Phi_0/2$ in the Wigner representation.
  The interaction of the SQUID loop with the environment (here modeled
  as a monochromatic thermal bath) locally destroys the interference
  between the two ``classical'' flux states $\ket{L}$ and $\ket{R}$
  represented by the localized peaks on either side. Figure reprinted
  with permission from \cite{Everitt:2004:mb}. Copyright
  2004 by the American Physical Society.  }
\end{figure}

Fig.~\ref{fig:squid-dec} illustrates this gradual disappearance of
interference in the symmetric ground state $\ket{\Psi_s} =
\frac{1}{\sqrt{2}} \bigl( \ket{L} + \ket{R} \bigr)$ due to the
interaction of the SQUID ring with a dissipative thermal bath in the
Wigner representation of the local density operator of the SQUID
\cite{Everitt:2004:mb} (see also \cite{Chudnovsky:2003:un}). As
predicted by the stability criterion, the robust states (i.e., the
preferred basis) selected by the environment are the macroscopically
distinguishable current states $\ket{L}$ and $\ket{R}$. The resulting
local loss of coherence---that is, the distribution of coherence,
initially associated with the SQUID, over the many degrees of freedom
of the SQUID-environment combination---constitutes the main obstacle
in the observation of coherent quantum tunneling.

\subsection{Molecular matter-wave interferometry} \label{sec:mol-interference}

\begin{figure}
\begin{center}
\includegraphics[scale=.21]{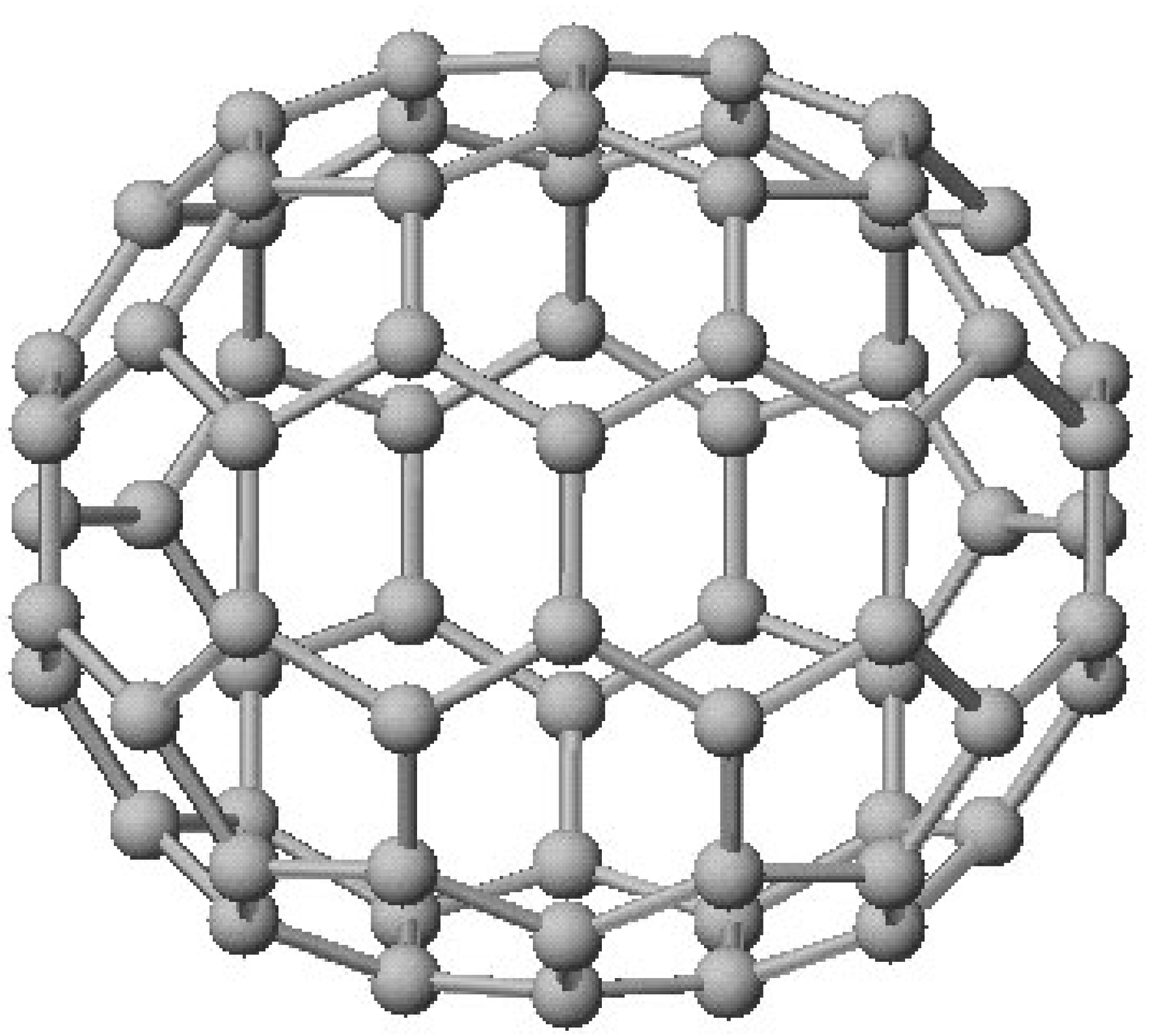} 
\includegraphics[scale=.20]{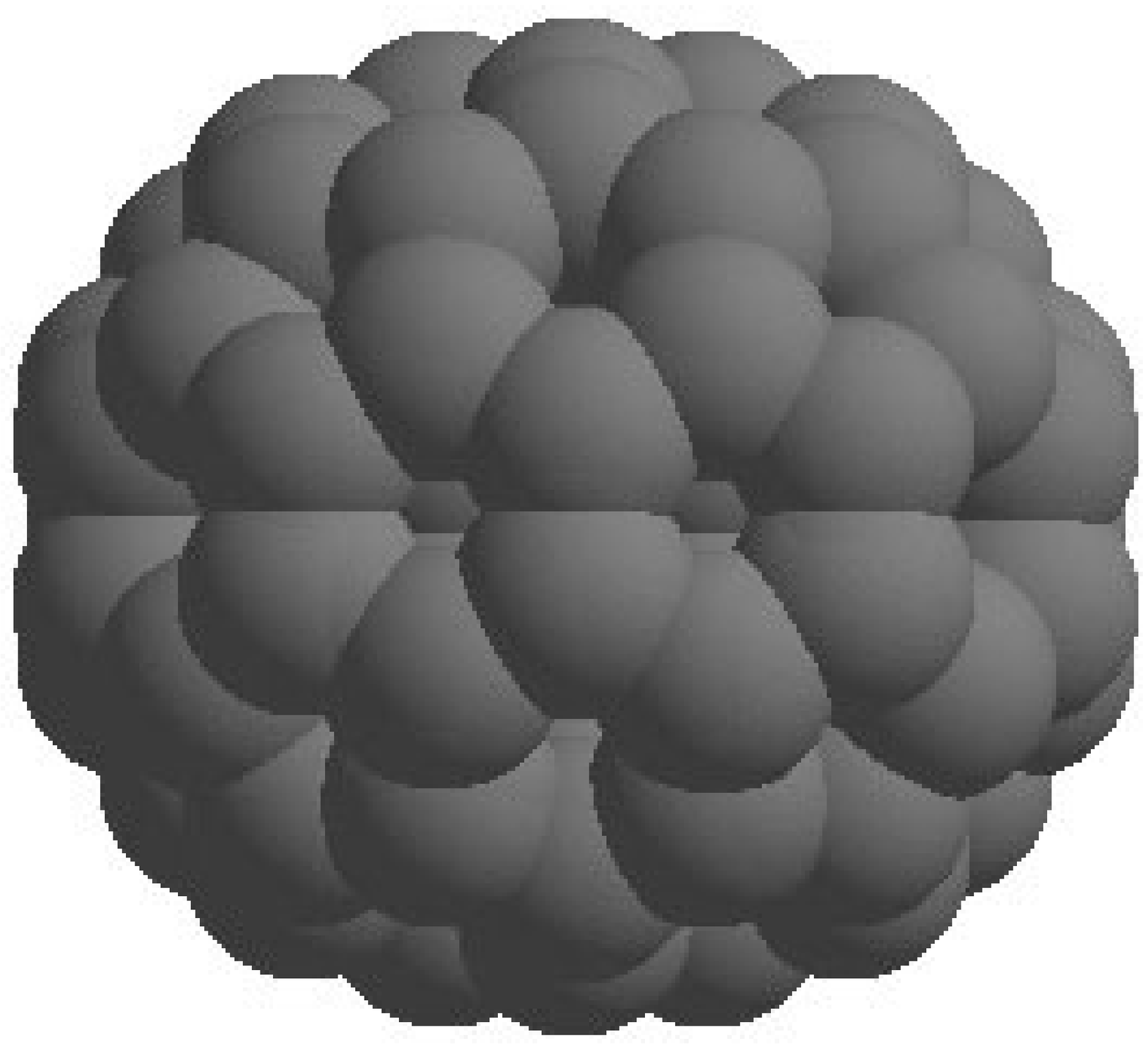}
\end{center}
\caption[Illustration of a C$_{70}$ molecule]{\label{fig:c70}
  Illustration of a C$_{70}$ molecule. The left image shows the
  ``backbone'' structure of interlinked carbon atoms. The right image
  displays the carbon atoms as massive spheres. }
\end{figure}

Recent experiments by the group of Zeilinger \etal
\cite{Arndt:1999:rc,Brezger:2002:mu,Hackermueller:2002:wb,Arndt:2002:bo,Nairz:2003:um,Hornberger:2003:tv,%
Hackermuller:2003:uu,Hackermuller:2004:rd,Hornberger:2005:mo} have pushed the boundary for the
observation of quantum (``wave'') behavior towards larger and larger
particles. In the experiment to be described, mesoscopic C$_{60}$
molecules (so-called fullerenes) and C$_{70}$ molecules have been
observed to lead to an interference pattern following passage through
a diffraction grating (``matter-wave interferometry'').  The carbon
atoms in the C$_{70}$ molecule are arranged in the shape of an
elongated buckyball with a diameter of about 1~nm (see
Fig.~\ref{fig:c70}). They are complex and massive enough to exhibit
properties that position them in the realm of classical solid objects
rather than that of atoms. For example, they possess a large number of
highly excited internal rotational and vibrational degrees of freedom
that allow one to attribute a finite temperature to each individual
molecule, and heated C$_{70}$ molecules are observed to emit photons
and electrons. The particle aspect seems to be overwhelmingly clear,
and yet these molecules have been shown to exhibit quantum
interference effects.

\begin{figure}
\begin{center}
\includegraphics[scale=.42]{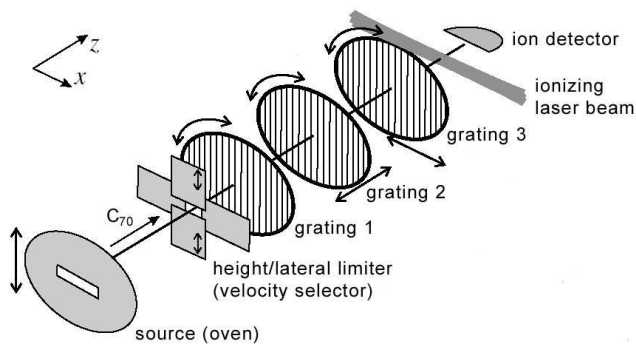}
\end{center}
\caption[Schematic sketch of a Talbot-Lau
interferometer]{\label{fig:talbotlau} Schematic sketch of a Talbot-Lau
  interferometer used for demonstrating quantum behavior of mesoscopic
  C$_{70}$ molecules. The first grating induces a certain degree of
  coherence in the incident uncollimated beam of molecules. The second
  grating then acts as the actual diffraction stage. Due to the
  Talbot-Lau effect, the molecular density pattern at the position of
  the third grating will be an image of the second grating if the
  molecules possess a quantum-wave nature. Scanning this pattern by
  moving the third grating (which acts as a mask) in the $x$-direction
  and detecting the transmitted and subsequently ionized molecules
  will then lead to an oscillatory signal that represents the
  interference effect.  Figure reprinted with permission from
  \cite{Brezger:2002:mu}. Copyright 2002 by the American Physical
  Society.}
\end{figure}

\subsubsection{Experimental setup and observation of interference}

The observation of C$_{70}$ interference patterns and their controlled
disappareance due to environmental decoherence induced by various
sources has been made possible by the so-called Talbot-Lau
interferometer \cite{Brezger:2002:mu} that has two main advantages
over earlier setups used for molecular interferometry
\cite{Borde:1994:om,Chapman:1995:bg}. First, the incident beam of
molecules does not need to be collimated, allowing for much higher
transmitted intensities.  Second, the required period of the gratings
used to obtain the interference pattern scales only with the square
root of the de~Broglie wavelength of the molecules, allowing for the
probing of the quantum behavior of, say, sixteen times larger
molecules by using an only four times smaller grating spacing.

The Talbot-Lau effect is based on the fact that the transverse part of
a plane wave $\psi(z) = \e^{ikz}$ incident on a periodic grating
located in the $xy$ plane will be identical to the grating pattern at
integer multiples of the distance (``Talbot length'')
\be
L_\lambda = \frac{d^2}{\lambda}
\ee
behind the grating. Since this is a pure interference effect, the
presence of the grating pattern at multiples of the Talbot length
indicates the wave nature of the incident beam.

The experimental setup that makes use of the Talbot-Lau effect is
shown schematically in Fig.~\ref{fig:talbotlau}. The main part
consists of a set of three gold gratings with a period of about
$d=1$~$\mu$m. The first grating acts as a collimator that induces a
sufficient degree of coherence in the incident uncollimated beam of
C$_{70}$ molecules in order to approximate the plane-wave assumption
made above. Each point of the grating can then be viewed as
representing a narrow source. The velocity of the molecules can be
selected over a range from about 80~m/s to 220~m/s, corresponding to
de~Broglie wavelengths of approximately 2--6~pm. The second grating is
the actual diffraction element, assuming the role of the single
grating in the above plane-wave example. The third grating, placed
behind the second grating at a distance $L$ equal to the Talbot length
$L_{\lambda_{\text{C}70}} = d^2/\lambda_{\text{C}70}$, where
$\lambda_{\text{C}70}$ is the de~Broglie wavelength of the molecules,
can be moved in the $x$-direction and serves as a scanning detection
mask for the molecular density pattern in the transverse plane at this
location.  The molecules that have passed through the third grating
are ionized by a laser beam and then counted by an ion detector.

\begin{figure}
\begin{center}
\includegraphics[scale=.37]{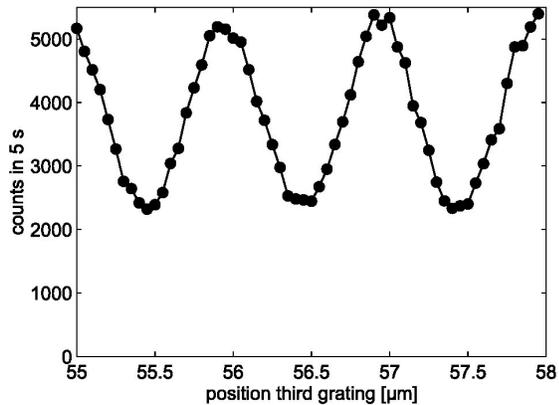}
\end{center}
\caption[Interference fringes for C$_{70}$
molecules]{\label{fig:c70-scan} Interference fringes for C$_{70}$
  molecules measured at the position of the third grating in a
  Talbot-Lau interferometer. Figure reprinted with permission from
  \cite{Brezger:2002:mu}. Copyright 2002 by the American Physical
  Society.}
\end{figure}

If the C$_{70}$ molecules indeed possess a quantum-wave nature, the
Talbot-Lau effect implies that the molecular density pattern at the
position of the third grating should consist of interference fringes
with a period equal to the spacing $d$ of the grating pattern. Thus,
when the third grating is scanned in the $x$-direction, we expect an
oscillation in the number of transmitted molecules with period $d$.
This is indeed what has been observed experimentally
\cite{Arndt:1999:rc,Arndt:2002:bo,Brezger:2002:mu,Nairz:2003:um,%
Hornberger:2003:tv} (Fig.~\ref{fig:c70-scan}).  The
possibility that these fringes could result from a classical blocking
of rays by the gratings (Moir\'e fringes) can be excluded, because such
patterns would be independent of the de~Broglie wavelength, in
contrast to what is observed experimentally
\cite{Brezger:2002:mu,Hackermuller:2003:uu}.  This confirms the
quantum origin of the measured fringes and thus the wave nature of the
C$_{70}$ molecules.

It should be emphasized that the fringes represent {\em
  single-particle} interference effects, rather than being due to
interference between different molecules \cite{Nairz:2003:um}. The
latter case would require the interfering molecules to be in the same
state, which is practically never the case due to the large number of
different excited internal states. Furthermore, the density in the
molecular beam is relatively low, such that the average distance
between two molecules is much greater than the range of any
intermolecular force. Thus, even if the molecules passed at such a
slow rate through the apparatus that only a single molecule was
present at any time, an interference pattern would emerge. The
interference effect is entirely due to the splitting and overlapping
of the wave fronts associated with each individual C$_{70}$ molecule.
This demonstrates clearly that quantum-mechanical superpositions in
configuration space describe individual states that can exhibit
interference effects (i.e., phase dependencies) without any
statistical aspect.

\subsubsection{Disappearance of interference
  due to controlled decoherence} \label{sec:diffrac-dec}

General numerical estimates for decoherence rates derived from
theoretical expressions
\cite{Joos:1985:iu,Gallis:1990:un,Tegmark:1993:uz,Hornberger:2003:un}
have clearly demonstrated the extreme efficiency of decoherence on
mesoscopic and macrosopic scales. It is therefore usually practically
impossible to control the environment in such a way as to explicitely
resolve and observe the gradual action of decoherence on larger
objects.

The Talbot-Lau interferometer, however, has made such observations
possible and has also led to direct confirmations of the predictions
of decoherence theory for mesoscopic objects
\cite{Hackermuller:2003:uu,Hornberger:2003:tv,%
Hackermuller:2004:rd,Hornberger:2004:bb,Hornberger:2005:mo}. The main sources of
decoherence that have been experimentally investigated are collisions
with gas molecules present in the interferometer
\cite{Hornberger:2003:tv,Hackermuller:2003:uu,Hornberger:2004:bb}, and
thermal emission of radiation when the C$_{70}$ molecules are heated
to temperatures beyond 1,000~K
\cite{Hackermuller:2004:rd,Hornberger:2005:mo}. Here, we shall focus
on the first case of decoherence, as collisions with environmental
particles represent the most natural and ubiquituous source of
decoherence in nature.

\begin{figure}
\begin{center}
  \includegraphics[scale=.37]{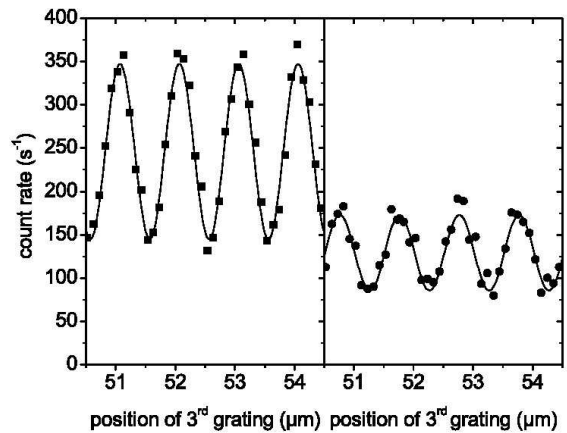}

\vspace{.5cm}

  \includegraphics[scale=.37]{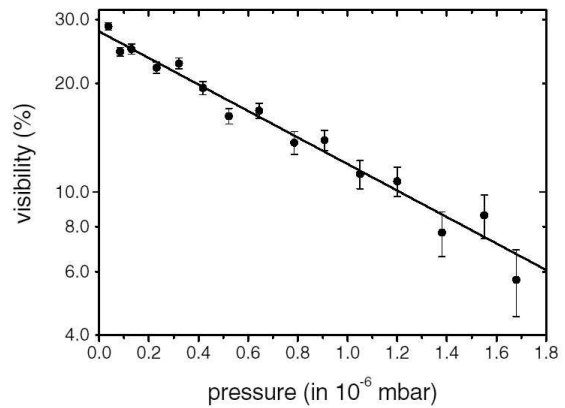}
\end{center}
\caption[Diminished interference effect in
C$_{70}$-molecule interferometry due to collisional
decoherence]{\label{fig:c70-dec} Diminished interference effect in
  C$_{70}$-molecule interferometry due to decoherence induced by
  collisions with gas molecules. {\em Above:} Decreased visibility of
  the interference fringes when the pressure of the gas is increased
  from the left to the right panel. {\em Below:} Dependence of the
  visibility on the gas pressure. The measured values (circles) are
  seen to agree well with predictions obtained from decoherence theory
  (solid line), see Eqs.~\eqref{eq:vis1} and \eqref{eq:vis2}. Figures
  reprinted, with kind permission of Springer Science and Business
  Media, from \cite{Hackermuller:2003:uu}.}
\end{figure}

In the experiments, the vacuum chamber containing the interferometer
is filled with gases at different pressures. Each collision between a
gas particle and a C$_{70}$ molecule entangles their motional states.
Since the C$_{70}$ molecules are much more massive than the gas
molecules, the motional state of the gas molecule is distinguishably
changed in the collision, while the motion of the C$_{70}$ molecule
remains essentially unaffected and can therefore still be detected at
the third grating. Thus, each collision encodes which-path information
about the trajectory of the C$_{70}$ molecule in the environment
(i.e., in the colliding gas particle). This leads to decoherence in
the spatial wave function of the C$_{70}$ molecules, since the
post-collision environmental states are approximately orthogonal in
the position basis due to the significant change of the motional state
of the gas molecules in the collisions.

To see this more explicitely, let us denote the state of the C$_{70}$
molecule before and after the scattering by 
\begin{align}
\ket{\psi}_{\text{C}70} &=
\int \, d\mathbf{x} \bigl( \braket{\mathbf{x}}{\psi}\bigr)_{\text{C}70}
\ket{\mathbf{x}}_{\text{C}70}  \\
  \intertext{and}  \ket{\psi'}_{\text{C}70} &= \int \,
d\mathbf{x} \bigl(\braket{\mathbf{x}}{\psi'}\bigr)_{\text{C}70}
\ket{\mathbf{x}}_{\text{C}70},
\end{align}
respectively, where 
\be
\bigl( \braket{\mathbf{x}}{\psi}\bigr)_{\text{C}70} \approx \bigl(
\braket{\mathbf{x}}{\psi'}\bigr)_{\text{C}70}
\ee
for all $\mathbf{x}$.  A collision at $\mathbf{X}$ changes the state
of the colliding gas molecule from $\ket{\varphi}_\text{gas}$ to
$\ket{\varphi', \mathbf{X}}_\text{gas}$, which encodes which-path
information about the C$_{70}$ molecule. Since the $\ket{\varphi',
  \mathbf{X}}_\text{gas}$ represent distinguishable motional states,
the environmental states corresponding to scattering events at
different locations become approximately orthogonal, 
\be
\bigl(
\braket{\varphi', \mathbf{X}}{\varphi', \mathbf{Y}}\bigr)_\text{gas}
\approx \delta(\mathbf{X} - \mathbf{Y}).
\ee
The collision leads to an entangled state for the combined
gas-C$_{70}$ system,
\begin{multline}
\ket{\Psi_0} = \ket{\psi}_{\text{C}70} \otimes
\ket{\varphi}_\text{gas} \\ \,\longrightarrow \,
\ket{\Psi} \approx  \int \, d\mathbf{X} \bigl(
\braket{\mathbf{X}}{\psi}\bigr)_{\text{C}70} 
\ket{\mathbf{X}}_{\text{C}70} \otimes \ket{\varphi', \mathbf{X}}_\text{gas}.
\end{multline}
The reduced density matrix for the C$_{70}$ molecule expressed in the
position basis is then obtained by averaging over all possible states
$\ket{\varphi', \mathbf{X}}_\text{gas}$ of the gas molecule,
\bea
\rho_{\text{C}70} &\approx& \int d\mathbf{X} \int d\mathbf{X}' \int
d\mathbf{X}'' \, \bigl(\braket{\mathbf{X}}{\psi}\bigr)_{\text{C}70}
\bigl(\braket{\mathbf{X}'}{\psi}\bigr)_{\text{C}70}^* 
\nonumber \\ && \times \, \bigl(\braket{\varphi',
  \mathbf{X}''}{\varphi', \mathbf{X}}\bigr)_\text{gas}  
\nonumber \\ && \times \, \bigl(\braket{\varphi', \mathbf{X}'}{\varphi',
  \mathbf{X}''}\bigr)_\text{gas} 
\bigl( \ketbra{\mathbf{X}}{\mathbf{X}'}\bigr)_{\text{C}70}  \nonumber \\
&\approx& \int d\mathbf{X} \, \bigl|
\bigl(\braket{\mathbf{X}}{\psi}\bigr)_{\text{C}70} \bigr|^2 \,
\bigl( \ketbra{\mathbf{X}}{\mathbf{X}}\bigr)_{\text{C}70},
\eea
where the vanishing of interference terms
$\bigl(\braket{\mathbf{X}}{\psi}\bigr)_{\text{C}70}
\bigl(\braket{\mathbf{X}'}{\psi}\bigr)_{\text{C}70}^*$, $\mathbf{X}
\not= \mathbf{X}'$, in the last step follows from the approximate
orthogonality of the $\ket{\varphi', \mathbf{X}}_\text{gas} $.  Thus,
the gas molecules carry away which-path information, leading to a
diffusion of coherence into the environment. Incidentally, in this
sense, Bohr's complementarity principle can be understood as a
consequence of entanglement: The observability of an interference
pattern, and thus the degree of the ``wave aspect'' of the C$_{70}$
molecules, is directly related to the amount of information, encoded
through entanglement with the state of the gas particles, about the
path (the ``particle aspect'') of the molecules.

We expect the visibility $V_\lambda$ of the interference fringes
(defined as $(c_\text{max} - c_\text{min}) / (c_\text{max} +
c_\text{min})$, where $c_\text{max}$ and $c_\text{min}$ are the
maximum and minimum amplitudes of the interference pattern) to
decrease as the pressure of the environmental gas is increased. A
theoretical analysis
\cite{Hackermuller:2003:uu,Hornberger:2003:un,Hornberger:2004:bb}
predicts that $V_\lambda$ will decrease exponentially with the
pressure $p = n k_B T$ of the colliding gas,
\be \label{eq:vis1}
V_\lambda(p) = V_\lambda(0) \e^{-p/p_0}.
\ee
Here, 
\be \label{eq:vis2}
p_0 = \frac{k_B T}{2L \sigma_\text{eff}}
\ee
is the characteristic decoherence constant (``decoherence pressure''),
where $L$ denotes the distance between the gratings and
$\sigma_\text{eff}$ corresponds to the effective cross section
\cite{Hackermuller:2003:uu}. This pressure-dependent decay of the
visibility has indeed been confirmed experimentally for C$_{70}$
molecules \cite{Hackermuller:2003:uu,Hornberger:2003:tv}, in excellent
agreement with the theoretical predictions (Fig.~\ref{fig:c70-dec}).

Studies of collision-induced decoherence in a Talbot-Lau
interferometer not only represent an outstanding method to observe the
gradual disappearance of quantum-interference effects while having
full control over both the source and the strength of decoherence, but
also allow one to predict the environmental conditions (in this case,
the maximum pressure of the surrounding gas) required to observe
quantum effects for even more complex and massive objects than tested
thus far. Such experiments are limited by two main factors
\cite{Hornberger:2003:tv,Hackermuller:2003:uu}. First, the velocity of
the objects must be quite slow during the passage through the
interferometer, in order to keep the de~Broglie wavelengths long
enough to allow for a sufficient degree of diffraction by practically
realizable gratings. Second, the pressure $p$ of the residual gas in
the interferometer must be low enough to maintain sufficient
visibility of the interference pattern, i.e., we must have $O(p) =
p_0$, see Eq.~\eqref{eq:vis2}.  Since both limits are purely technical
and can be precisely quantified, there is no indication for any fixed
quantum-classical boundary in this case other than the observational
limit determined by environmental decoherence, for which rigorous
theoretical estimates can be given. Decoherence allows for an exact
specification of where the quantum-to-classical transition occurs and
what needs to be done to move the boundary.

\begin{figure}
\begin{center}
  \includegraphics[scale=.43]{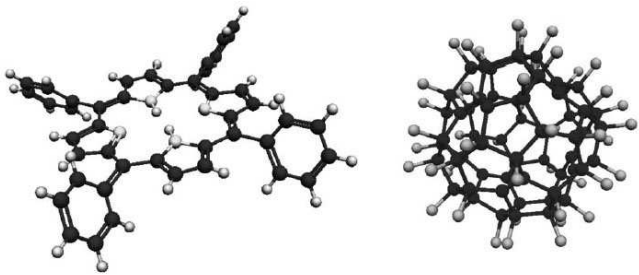}
\end{center}
\caption[Structure of the biomolecule
tetraphenylporphyrin C$_{44}$H$_{30}$N$_4$ and of the fluorofullerene
C$_{60}$F$_{48}$]{\label{fig:biomol} Structure of the biomolecule
  tetraphenylporphyrin C$_{44}$H$_{30}$N$_4$ (left) and of the
  fluorofullerene C$_{60}$F$_{48}$ (right). The wave nature of both
  molecules has been observed in experiments.  Figures reprinted with
  permission from \cite{Hackermueller:2002:wb}.  Copyright 2003
  by the American Physical Society.}
\end{figure}

\begin{figure}
\begin{center}
  \includegraphics[scale=.35]{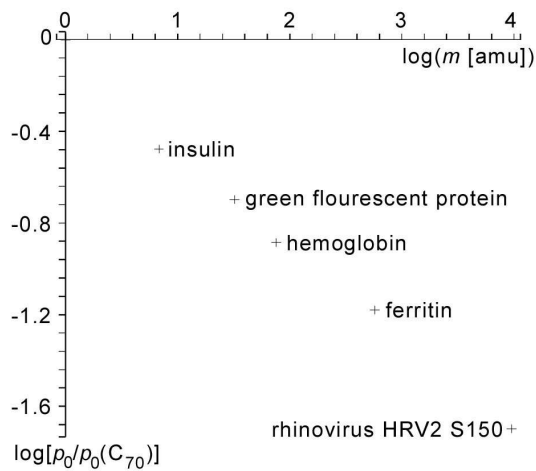}
\end{center}
\caption[Extrapolated maximum residual air pressures for 
interference experiments with various
biomolecules]{\label{fig:extrapol}Extrapolated maximum residual air
  pressures $p_0$, relative to the value of $p_0$ for C$_{70}$
  molecules, versus molecular weight $m$ (in amu), that would allow
  for an observation of interference fringes for various biological
  structures in an elongated ($L = 1$~m) Talbot-Lau interferometer.
  Data from \cite{Hackermuller:2003:uu}.  }
\end{figure}

In fact, the envelope for the observation of the wave nature of
mesoscopic molecules has recently been pushed even further in
experiments demonstrating quantum interference fringes for the
important biomolecule tetraphenylporphyrin C$_{44}$H$_{30}$N$_4$ (with
mass $m = 614$~amu and a width over 2~nm) and for the fluorinated
fullerene C$_{60}$F$_{48}$ (mass $m = 1632$~amu, 108~atoms)
\cite{Hackermueller:2002:wb}. While tetraphenylporphyrin is the
first-ever biomolecule whose wave nature has been demonstrated
experimentally, fluorofullerenes are the most massive and complex
molecules to exhibit quantum behavior thus far. Theoretical estimates
for the maximum residual gas pressure that would still allow for the
observation of interference fringes for even larger biological
objects, up to the size of a rhinovirus, have been given by
Hackerm\"uller \etal \cite{Arndt:2002:bo,Hackermuller:2003:uu} (see
Fig.~\ref{fig:extrapol}) and appear to be realizable even with the
currently available technology in Talbot-Lau interferometry
\cite{Hornberger:2003:tv,Hackermuller:2003:uu}. One might extrapolate
even further and speculate about the feasibility of interference
experiments involving human cells, with an average weight and size on
the order of $10^{15}$~amu and $10^4$~nm, respectively.  While this is
certainly beyond the existing technology, there is no reason to assume
that such experiments should be impossible.

\subsubsection{Implications of the C$_{70}$ interference experiments}

The described matter-wave interferometry experiments have led to three
crucial results:

\bn

\item Interference patterns are observed for particles that clearly
  reside in the ``lump of matter'' category.
  
\item These patterns are due to single-particle (rather than
  interparticle) interference effects.
  
\item Any observed disappearance (or absence) of interference patterns
  can be well understood as resulting from decoherence and can be
  explicitly controlled and quantified.

\en

Thus there is no theoretical or experimental indication for any
fundamental limit on the ability of objects to exhibit quantum
behavior (i.e., a wave nature) if these objects are sufficiently
shielded from the decohering influence of their environment. Result
(2) shows that the initial wave function describing the individual
molecule evolves into a spatially extended wave function after passage
through the diffraction grating, namely, into a superposition of
``classical'' localized position states that each correspond to the
molecule being in a specific region of space. The gradual
disappearance of interference due to controlled interaction with the
environment can be understood as entanglement between the different
relative states of the environment and the individual components
$\ket{\mathbf{x}}_{\text{C}70}$ in the superposition. It is important
to note that all components $\ket{\mathbf{x}}_{\text{C}70}$ are still
present regardless of the environmental interaction --- decoherence is
in principle fully reversible, as experiments on coherent state-vector
revival have shown (see, e.g., \cite{Raimond:1997:um}).

\subsection{Bose-Einstein condensation} \label{sec:bose}

\begin{figure}
\begin{center}
  \includegraphics[scale=.218]{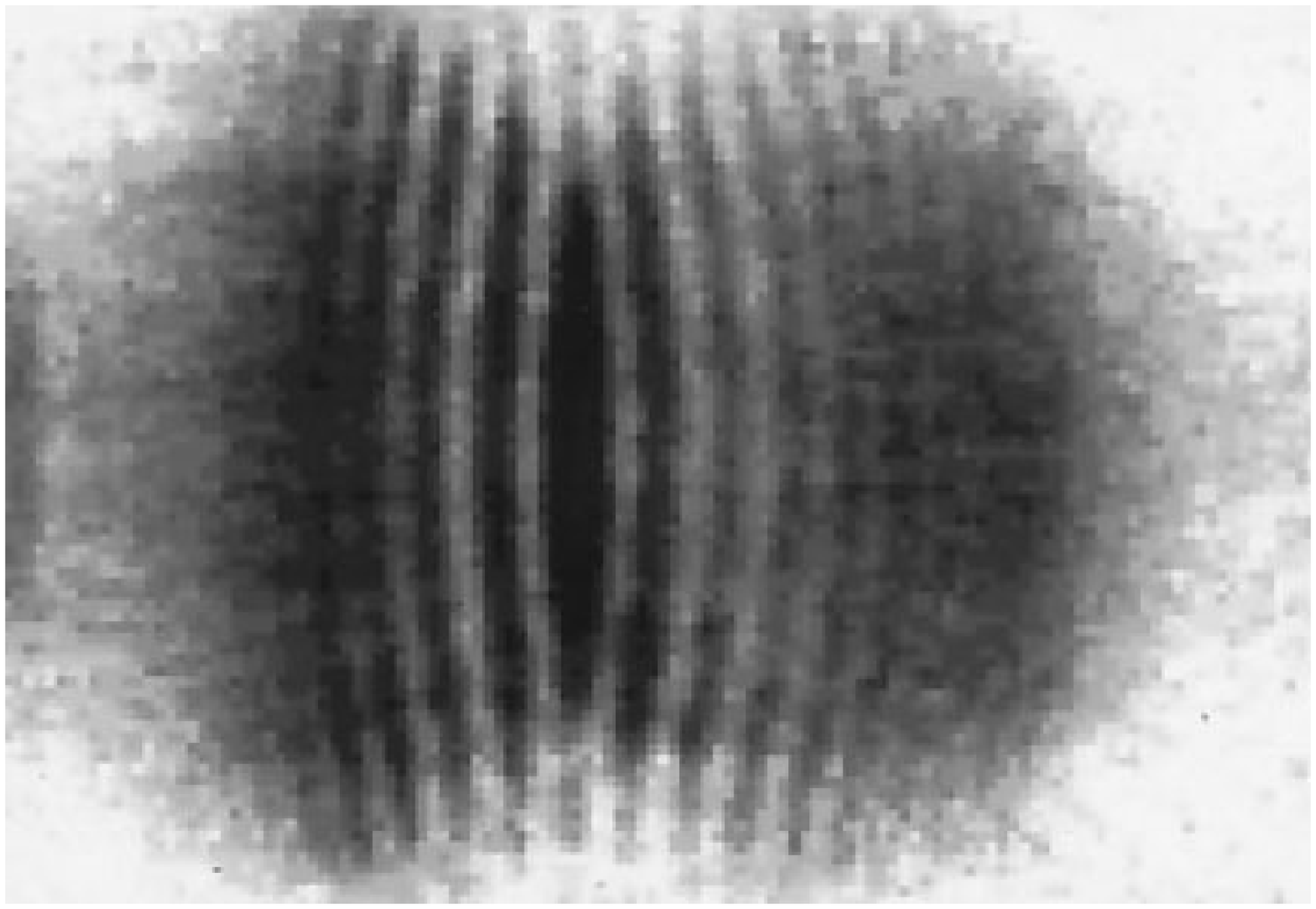}
  \includegraphics[scale=.2]{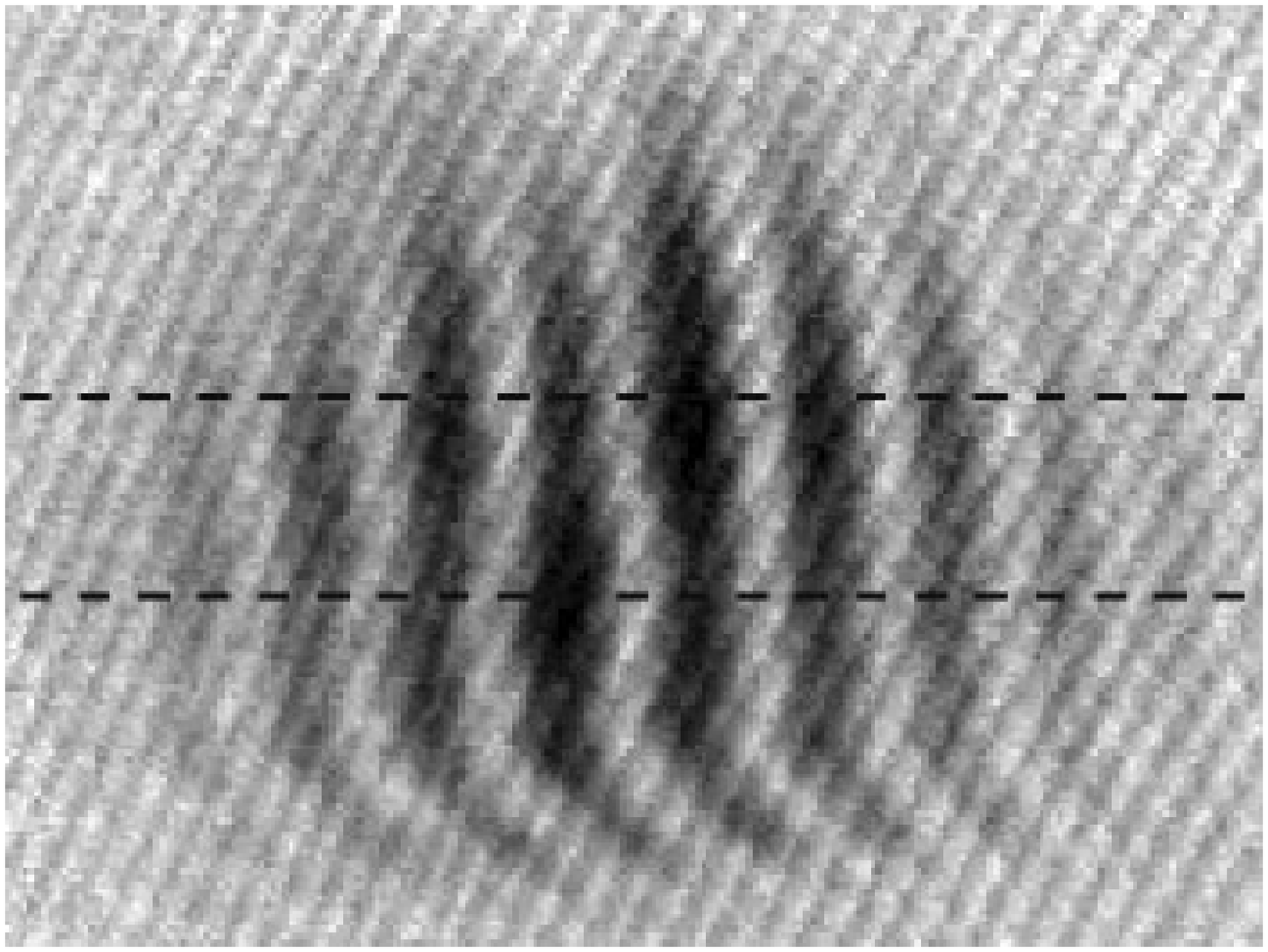}
\end{center}
\caption[Matter-wave interference pattern for the atomic gas
density in Bose-Einstein condensates]{\label{fig:bec1} Matter-wave
  interference pattern for the atomic gas density in Bose-Einstein
  condensates. {\it Left:} Pattern obtained by letting two independent
  condensates overlap, demonstrating that the condensate is indeed
  described by a single wave function with a phase. The fringe period
  was measured to be 20~$\mu$m. Figure reprinted with permission
  \cite{Andrews:1997:um}. Copyright 1997 by AAAS. {\it Right:}
  Fringes due to interference of a single coherently split condensate.
  This experiment corresponds to a BEC ``double-slit'' interferometer.
  Figure reprinted with permission from \cite{Shin:2004:lo}.
  Copyright 2004 by the American Physical Society.  }
\end{figure}

As a third example, we shall discuss Bose-Einstein condensation (BEC).
While this effect had been predicted theoretically already in the
1920s by Einstein
\cite{Einstein:1924:om,Einstein:1925:om,Einstein:1925:oy} based on
ideas by Bose \cite{Bose:1924:om}, explicit experimental verification
succeeded only in 1995
\cite{Bradley:1995:om,Davis:1995:mm,Anderson:1995:mx,Bradley:1997:nu}.
When an atomic bosonic gas confined by a magnetic trap is cooled down
to very low temperatures, the de~Broglie wavelength $\lambda_\text{dB}
= (2\pi\hbar^2/mk_BT)^{1/2}$ associated with each atom becomes long in
comparison with the interparticle separation. At a precise temperature
in the $\approx 100$~nK range, the collection of atoms can undergo a
quantum-mechanical phase transition to a condensate in which the atoms
lose their individuality and all occupy the same quantum state. Then a
macrosopic number of atoms---large condensates can contain of the
order of $10^7$ atoms---is described by a single $N$-particle wave
function with a phase,
\be
\Psi_N(\mathbf{r}_1, \mathbf{r}_1, \cdots
,\mathbf{r}_N) = \e^{\mathrm{i}\Phi} \prod_{i=1}^N |\psi(\mathbf{r}_i)|, 
\ee
i.e., as a product of $N$ identical single-particle wave functions
$\psi(\mathbf{r})$. As a consequence, BECs can directly exhibit
quantum behavior. For instance, two condensates released from adjacent
traps can overlap and form a gas-density interference pattern due to
the phase difference between the two wave functions
(Fig.~\ref{fig:bec1})
\cite{Javanainen:1996:mm,Andrews:1997:um,Rohrl:1997:pp,Javanainen:2005:mo,Saba:2005:lm}.
Recently, Bose-Einstein ``double-slit'' interferometers have been
experimentally realized \cite{Shin:2004:lo} and theoretically analyzed
\cite{Collins:2005:uu}. Here, a single condensate is coherently split
(corresponding to the diffraction stage in the double-slit experiment)
and then allowed to recombine, which leads to the observation of
interference fringes (Fig.~\ref{fig:bec1}).

\subsubsection{Macrosopic number-difference superpositions using Bose-Einstein condensates}

Various methods have been proposed for the creation of BEC-based
Schr\"odinger cat states in form of a superposition of states with
macroscopically distinguishable numbers of particles
\cite{Cirac:1998:mm,Ruostekoski:1998:mm,Gordon:1999:mh,Dunningham:2001:da,%
Calsamiglia:2001:tt,Louis:2001:mu,Micheli:2003:jn}.  BECs are
particularly suitable for the generation and the study of
Schr\"odinger cat states, for several reasons. First, as BECs involve
up to $10^7$ atoms, such superpositions would be the most macrosopic
ones ever observed.  Second, the condensate is described by a single
coherent wave function that pertains to a controllable number of atoms
and possesses an extremely long coherence time (up to 10--20~s).
Third, the sources of decoherence (mostly loss of particles from the
condensate) are fairly well-understood and potentially sufficiently
controllable through suitable environmental engineering and trap
design \cite{Ruostekoski:1998:tt,Kuang:1999:tb,Dalvit:2000:bb}.

The typically suggested scheme to create superpositions of
macrosopically distinguishable states using BECs involves the creation
and manipulation of interacting two-species condensates, i.e., BECs in
which the atoms possess two different internal states $\ket{A}$ and
$\ket{B}$. Experimental realizations of two-species BECs often employ
the two hyperfine sublevels $\ket{F,m_F} = \ket{2,1}$ and $\ket{1,-1}$
of $^{87}$Rb. The early proposal by Cirac \etal \cite{Cirac:1998:mm}
(similiar models have been suggested, for
example, in \cite{Ruostekoski:1998:mm,Gordon:1999:mh,%
  Dunningham:2001:da,Micheli:2003:jn,Jack:2005:oo}) is based on a Josephson-like
coupling between the two species that leads to a number-difference
superposition of the form
\be \label{eq:bec-cat1}
\ket{\Psi} = \frac{1}{\sqrt{2}} \bigl( \ket{n_A,N-n_A} + \e^{\mathrm{i}\varphi}
\ket{N-n_A,n_A} \bigr),
\ee 
where $\ket{n_A, n_B}$ is the occupation-number state representing
$n_A$ atoms of type A and $n_B$ atoms of type B, and $N=n_A+n_B$ is
the total number of atoms. This represents a superposition of two
states which differ by a macrosopic number $|N-2n_A|$ of atoms of a
certain type (A or B). Then $n_A=0$ or $n_A=N$ would correspond to a
maximally entangled $N$-particle GHZ-type state
\cite{Greenberger:1990:bw} and thus the most ``cat-like'' state
\be \label{eq:bec-cat2}
\ket{\Psi} = \frac{1}{\sqrt{2}} \bigl( \ket{N,0} + \e^{\mathrm{i}\varphi}
\ket{0,N} \bigr).
\ee 
Another scheme for the creation of macrosopic BEC superpositions that
uses a single-component BEC in a double well (with possible
generalizations to $M$ wells) has been described in
\cite{Mahmud:2004:rz,Mahmud:2005:rz} (see also
\cite{Polkovnikov:2002:ii,Polkovnikov:2003:ll}).  Here, a
laser-induced phase shift is imprinted on the condensate in one of the
wells, followed by a change of barrier height. This is predicted to
lead to a superposition of the form
\be\label{eq:bec-cat3}
\ket{\Psi} = \frac{1}{\sqrt{2}} \bigl( \ket{n_L,N-n_L} + \e^{\mathrm{i}\varphi}
\ket{N-n_L,n_L} \bigr), 
\ee
where $\ket{n_L,n_R}$ is the number state corresponding to $n_L$
($n_R$) atoms in the left (right) well. Again, $n_L$ determines the
degree of entanglement, with $n_L=0$ or $n_L=N$ corresponding to
maximal ``catness.''  Even the possibility of creating a coherent
superposition of a macroscopic number of atoms with a macroscopic
number of molecules using photoassociation in BECs (i.e., the
absorption of a photon by two atoms, leading to the formation of a
two-atomic bound molecule) has been indicated
\cite{Calsamiglia:2001:tt}.

To detect a BEC cat state, one might in principle envision experiments
similiar to those measuring GHZ spin states
\cite{Mermin:1990:un,Dalvit:2000:bb,Sackett:2000:uu}, although this
would be very difficult to carry out in practice for the larger values
of $N$ relevant to BEC superpositions. Instead, as pointed out in
\cite{Dalvit:2000:bb}, one could first confirm that measurement
statistics indeed give equal likelihoods for the two cat-state terms
$\ket{N,0}$ or $\ket{0,N}$. If the system can also be observed to
(approximately) return to its initial state after unitary evolution
over a period that is an integer multiple of the time needed for the
generation of the cat state, this would provide strong indications for
the presence of a cat state.

\subsubsection{Decoherence of BEC superpositions}

To date, Schr\"odinger-cat states using BECs have not been realized
experimentally, although much progress has been made (see, for
example, \cite{Albiez:2004:nn}). Dissipation and decoherence
effects are still too strong to allow for a direct observation of
superpositions and will continue to constitute the dominant limit on
the size of number-difference Schr\"odinger cats. These environmental
effects are mainly due to elastic and inelastic scattering between
condensate and noncondensate atoms.

Elastic collisions with noncondensate atoms under conservation of the
number of condensate atoms lead to phase damping and thus to the
destruction of the coherent superposition. The reduced density matrix
in the number basis then decoheres according to \cite{Louis:2001:mu}
\be
\bra{m}\widehat{\rho}(t)\ket{n} =  \e^{-(m-n)^2
  \kappa t}  \bra{m}\widehat{\rho}(0)\ket{n} \e^{-\mathrm{i}\omega(m-n)t},
\ee
i.e., the off-diagonal elements $m\not= n$ decay with a decoherence
rate that scales with the square of the number difference, $(m-n)^2$.

Furthermore, inelastic collisions with noncondensate atoms lead to a
loss of atoms from the condensate, which diminishes coherence.  Again,
the larger the number difference in the superposition $\bigr(
\ket{n,N-n} + \ket{N-n,n}\bigr) / \sqrt{2}$ is (i.e., the closer $n$
is to $0$ or $N$), the more sensitive the state is to atom loss (see,
for example, the detailed analysis in \cite{Dunningham:2001:da}).
In the limit of the maximally entangled state $\bigl( \ket{N,0} +
\ket{0,N} \bigr) / \sqrt{2}$, already the loss of a single atom of,
say, type 1 completely destroys the coherent superposition, since
\be
\widehat{a}_1 \bigl( \ket{N,0} + \e^{\mathrm{i}\varphi} \ket{0,N} \bigr) / \sqrt{2} = \sqrt{N/2
} \, \ket{N-1,0},
\ee
where $\widehat{a}_1$ is the destruction operator for particles of
type~1. 

Thus both decoherence effects will usually limit the size $N$ (i.e.,
the number difference) of superpositions of the form $\bigl( \ket{N,0}
+ \ket{0,N} \bigr) / \sqrt{2}$. In a detailed analysis that combines
the two forms of scattering processes, Dalvit \etal \cite{Dalvit:2000:bb}
have estimated the decoherence rate $\tau_d^{-1}$ for an optimal
number-difference superposition $\bigl( \ket{N,0} + \ket{0,N} \bigr) /
\sqrt{2}$ in a standard harmonic trap due to a ``thermal cloud'' of
$N_\text{nc}$ noncondensate atoms as
\be
\tau_d^{-1} \propto  a^2 N_\text{nc} N^2,
\ee
where $a$ is the scattering length. This leads to very short
decoherence times even for moderate environment and condensate sizes
\cite{Dalvit:2000:bb,Louis:2001:mu}. For example, for $N_\text{nc}=10$
and $N=10^3$, $\tau_d$ is of the order of milliseconds. For larger
Schr\"odinger cats with $N=10^7$ and a thermal cloud containing
$N=10^4$ noncondensed atoms, $\tau_d \sim 10^{-13}$~s.  

However, several schemes exist to significantly reduce the decoherence
rate and to thus render it quite likely that BEC-based
number-difference Schr\"odinger cat states could indeed be observed in
future experiments; for example:

\bn

\item The construction of modified traps that allow for a faster
  evaporation of the thermal cloud \cite{Dalvit:2000:bb}.
  
\item Generation of number-difference cat states via the creation of
  macrosopic superpositions of relative-phase states that are not only
  much less sensitive to atom loss, but might even {\it require} such
  loss \cite{Dunningham:2001:da}.
  
\item A ``symmetrization'' of the environment to reduce the effective
  size of the thermal cloud \cite{Dalvit:2000:bb}.
  
\item Sufficiently fast generation of the cat state
  \cite{Micheli:2003:jn}.

\en

The key lesson to be learned from the example of BEC-based
Schr\"odinger-cat states is that, nonwithstanding the fact that such
superpositions have not (yet) been explicitly documented in
experiments, the physics of these states and the required conditions
to create them is very well understood. The failure to experimentally
generate these states with currently available setups is
well-explained by decoherence models that provide precise numerical
estimates for the type of experimental arrangements and parameter
ranges that would be required to observe Schr\"odinger-cat states
using BECs.  Similiar to the case of studying the feasibility of
matter-wave interferometry with larger molecules than those
investigated thus far (see Sec.~\ref{sec:mol-interference}),
decoherence is the key tool for a precise prediction of the physical
conditions required for the experimental observation of superpositions
of macrosopically distinct states.

\subsection{Analysis of the degree of
  macroscopicity of the experimentally achieved superpositions} \label{sec:scaling}

\begin{table}
\begin{center}
\begin{tabular}{lccc}
Experiment & $\mathcal{S}_\text{ext}$ & $\mathcal{S}_\text{ent}$ &
$\mathcal{S}_\text{ext} \times \mathcal{S}_\text{ent}$ \\ \hline
SQUID & $10^{10}$ & $10^9$ & $10^{19}$ \\
C$_{70}$ &  $10^6$ & $10^3$ & $10^9$ \\
Bose-Einstein$^*$ & $10^7$ & $10^9$ & $10^{16}$ \\ \hline
\multicolumn{4}{l}{$^*$not yet experimentally achieved}
 \\ 
\end{tabular}
\end{center}
\caption[Estimates for the degree of macrosopic
  distinctness of the states in superpositions]{
  Estimates for the degree of macrosopic 
  distinctness of the states in superpositions relevant to the three
  experiments discussed in this paper. $\mathcal{S}_\text{ext}$
  is a measure for the maximum difference in a suitably chosen
  extensive variable that distinguishes the states in the
  superposition (here: the total magnetic moment in SQUID experiments; the
  average separation between two paths in the interferometer in
  C$_{70}$ molecular diffraction; the difference in angular
  momentum in two-species Bose-Einstein
  condensates). $\mathcal{S}_\text{ent}$ measures the degree of
  entanglement in multi-particle states and is well-estimated by the
  number of microsopic constituents involved in the superposition
  (i.e., the  number of Cooper pairs in SQUID, and the number of nucleons and
  electrons in the C$_{70}$ molecule and the Bose-Einstein
  condensate). The third column shows the product
  $\mathcal{S}_\text{ext} \times \mathcal{S}_\text{ent}$ of the two
  measures, thus representing the overall degree of macroscopicity of the
  superpositions. See also Sec.~\ref{sec:measures}.} \label{tab:degrees}
\end{table}

In the following, let us compare the degree of macrosopic distinctness
of the states in the superpositions encountered in the experiments
with SQUIDs, diffracted molecules, and BECs.  We will use the
combination of the two measures introduced in Sec.~\ref{sec:measures},
namely, the difference $\mathcal{S}_\text{ext}$ in a relevant
extensive quantity between the states in the superposition relative to
an appropriate microsopic reference value, and the degree of
entanglement $\mathcal{S}_\text{ent}$ present in the multi-particle
superposition.

For the SQUID experiments (Sec.~\ref{sec:squid}), choosing the total
magnetic moment to be the relevant extensive variable, the two states
$\ket{L}$ and $\ket{R}$ differ by about $10^{10} \mu_B$ in the
experiment by Friedman \etal \cite{Friedman:2000:rr}. Taking the Bohr
magneton $\mu_B$ as the reference unit, the extensive difference
$\mathcal{S}_\text{ext}$ between the two states is thus of the order
of $10^{10}$. The degree of entanglement $\mathcal{S}_\text{ent}$ in
the multi-Cooper-pair state can be estimated to be of the order of the
number of Cooper pairs, i.e., $\sim 10^9$.

In the case of diffraction of C$_{70}$ molecules
(Sec.~\ref{sec:mol-interference}), a suitable extensive quantity would
be the center-of-mass displacement between the two paths through the
interferometer, which we can estimate to be on the order of 1~mm
(corresponding to the lateral width of the molecular beam
\cite{Hackermuller:2003:uu}) relative to the size of the molecule of
about 1~nm, which yields a value for $\mathcal{S}_\text{ext}$ on the
order of $10^6$. The degree of entanglement $\mathcal{S}_\text{ent}$
is essentially given by the number of microsopic constituents in the
molecule, $3 \times 6 \times 70 \sim 10^3$.

For BEC two-species superpositions that use the two hyperfine
sublevels $\ket{F,m_F} = \ket{2,1}$ and $\ket{1,-1}$ of $^{87}$Rb
atoms (Sec.~\ref{sec:bose}), a suitable extensive variable would be
the total difference in angular momentum due to the hyperfine
splitting, in units of $\hbar$, which is on the order of the number
$N$ of atoms in the condensate, which can be as large as $10^7$. Thus
the maximum $\mathcal{S}_\text{ext}$ is on the order of $10^7$. The
degree of entanglement $\mathcal{S}_\text{ent}$ is again suitably
measured by the number of nucleons and electrons in the condensate,
which is of the order of $100N$ for $^{87}$Rb. Note, however, that
such superpositions have not yet been experimentally achieved.
   
All values are summarized in Table~\ref{tab:degrees}. We see that the
SQUID experiments allow for superpositions that are about ten orders
of magnitude ``more macrosopic'' (in the sense defined above) than
those achieved by molecular interferometry. On the other hand, the
latter experiments lead to a direct realization of spatial
superpositions, which are often considered to be more
``counterintuitive'' than the superposition of superconducting
currents, since position appears to be the dominant definite quantity
in our observation of the macroworld. The ubiquitous perception of
definiteness in position space has even led some to postulate a
fundamentally preferred role to position. For example, Bell
\cite{Bell:1982:ag} stated that ``in physics the only observations we
must consider are position observations, if only the positions of
instrument pointers.'' A similiar idea underlies the spatial
localization mechanism in the GRW theory and is reflected in the
concept of definite particle trajectories in Bohmian mechanics.

Superpositions involving two-species BECs, if experimentally realized,
would come close to the degree of macroscopicity achieved in SQUIDs.
This result can be understood by noting the striking analogies between
the two experiments. In both cases, the multi-particle system (the
superconducting material in SQUIDs, or the atomic gas in BECs) is
cooled down to extremely low temperatures near absolute zero. The two
macrosopically distinguishable states (currents of opposite direction
in SQUIDs, or different atom species in BECs) are coupled by a
classically impenetrable barrier of the Josephson-junction type. In
both experiments, this essentially leads to Schr\"odinger-cat states
of the form
\be 
\ket{\Psi} = \frac{1}{\sqrt{2}} \bigl( \ket{N,0} + \e^{\mathrm{i}\varphi}
\ket{0,N} \bigr),
\ee 
where the number state $\ket{N,0}$ denotes $N$ particles (Cooper pairs
in SQUIDs, or atoms in BECs) being in the first macroscopically
distinguishable state (representing a clockwise current in the SQUID,
or the hyperfine sublevel $\ket{F,m_F} = \ket{2,1}$ in BECs), and no
particles being in the second state (corresponding to a
counterclockwise current in the SQUID, or the hyperfine sublevel
$\ket{F,m_F} = \ket{1,-1}$ in BECs).

\section{The status of physical collapse models}  \label{sec:physcoll}

All existing interpretations of quantum mechanics can be viewed as
either adding formal rules\footnote{As, for example, done in the
  Copenhagen interpretation (that formally postulates a collapse, but
  regards it merely as an ``increase of information,'' rather than as
  a physical process, since it interprets the wave function as
  representing a probability amplitude), Bohmian mechanics, modal
  interpretations, and consistent-histories interpretations.} or
physical elements (as in collapse models) to the axioms of minimal
quantum theory stated in the Introduction. With respect to the
``formal'' category, if the minimal theory can be shown to be
sufficient to explain and predict all our observations, there is
clearly no empirical reason for introducing purely formal additives.
While a similiar argument can be made regarding the ``physical''
category, collapse theories might lead to observable deviations from
Schr\"odinger dynamics and could thus be experimentally tested.  In
both cases, of course, there may be conceptual reasons that motivate
the added elements, for example a desire to resolve a felt
``weirdness'' in the existing quantum theory. While we respect this
motivation, we hope to show that in fact the minimal theory is
sufficient to resolve the problems without requiring any such
additions.

The increasing size of physical systems for which interference effects
have been observed imposes bounds on the parameters used in collapse
models.  However, the current experiments demonstrating mesoscopic and
macrosopic interference are still quite far away from disproving the
existing collapse theories. For example, even the C$_{70}$
diffraction experiments described in Sec.~\ref{sec:experiments} still
fall short of ruling out continuous spontaneous localization models
\cite{Pearle:1989:cs,Diosi:1989:yb,Ghirardi:1990:lm} (which lead to
the strongest deviations from Schr\"odinger dynamics among all
physical collapse theories) by eleven orders of magnitude
\cite{Adler:2004:om}. A recently proposed mirror-superposition
experiment by Marshall \etal \cite{Marshall:2003:om} that might lead
to a superposition involving of the order of $10^{14}$ atoms still
fails to rule out continuous spontaneous localization models by about
six orders of magnitude \cite{Bassi:2005:om}. The superpositions
observed in coherent quantum tunneling in SQUIDs also appear to be
compatible with dynamical reduction models, since the spatial
localization mechanism would only result in a small reduction of the
supercurrent below the detectable level due to a breaking-up of Cooper
pairs, but not in an approximate reduction onto one of the current
states \cite{Rae:1990:wa,Buffa:1995:mo,Bassi:2003:yb}. However, given
the rapid development of experiments that propose to demonstrate
quantum superpositions on increasingly large scales, it appears to be
only a matter of time to probe the range relevant to a test of
physical reduction models.

It is important to note that no deviations from linear Schr\"odinger
dynamics have ever been observed that could not also be explained (at
least in principle) as apparent deviations due to decoherence. In
fact, it would be very difficult to distinguish collapse effects from
decoherence, since the large number of atoms required for the collapse
mechanism to be effective also leads to strong decoherence
\cite{Tegmark:1993:uz,Benatti:1995:re,Bassi:2003:yb}. It would
therefore be necessary to isolate the system of interest extremely
extremely well from its environment, such that decoherence effects can
be neglected with respect to the environment-independent localization
mechanism. Even in this case it might be difficult to exclude the
influence of decoherence due to, for example, thermal emission of
radiation, as demonstrated in the case of fullerene and C$_{70}$
interferometry \cite{Hackermuller:2004:rd,Hornberger:2005:mo}.
  
This leaves physical collapse theories, at least so far, in the
speculative realm, with the added difficulty of obtaining relativistic
generalizations \cite{Bassi:2003:yb}.  Certainly, such collapse
mechanisms might be discovered in the future. However, in the absence
of positive experimental evidence for such effects, and given the
viable option of constructing a quantum theory consistent with all
observations from the minimal formalism alone (a strategy advocated in
this paper), the need for a postulated collapse effect, with free
parameters tuned such as to avoid inconsistencies with the observation
(or nonobservation) of superpositions, appears rather doubtful.

\section{Emergence of probabilities in a relative-state framework} \label{sec:prob}

The question of the origin and meaning of probabilities in a relative
state--type interpretation that is based solely on a deterministically
evolving global quantum state, and the problem of how to consistently
derive Born's rule in such a framework, has been the subject of much
discussion and criticism aimed at this type of interpretation (see,
e.g., \cite{Kent:1990:nm}). Several decoherence-unrelated
proposals have been put forward in the past to elucidate the meaning
of probabilities and to arrive at the Born rule in an explicit or
implicit relative-state context (see, for
instance, \cite{Everett:1957:rw,Hartle:1968:gg,DeWitt:1971:pz,%
  Graham:1973:ww,Geroch:1984:yt,Deutsch:1999:tz}). However, it is
highly controversial whether these approaches are successful and
represent a noncircular derivation
\cite{Stein:1984:uu,Kent:1990:nm,Squires:1990:lz}. A derivation that
is only based on the non-probabilistic axioms of quantum mechanics and
on elements of classical decision theory has been presented by Deutsch
\cite{Deutsch:1999:tz}. It was critized by Barnum \etal
\cite{Barnum:2000:oz}, but was subsequently defended by other authors
\cite{Gill:2003:tz,Wallace:2003:zr} and embedded into an operational
framework by Saunders \cite{Saunders:2002:tz}. It is fair to say that
no decisive conclusion appears to have been reached as to the success
of these derivations.

Initially, decoherence was thought to provide a natural account of the
probability concept in a relative-state framework, by relating the
diagonal elements of the decohered reduced density matrix to a
collection of possible ``events'' that can be reidentified over time,
and by interpreting the corresponding coefficients as relative
frequencies of branches, thus leading to an interpretation of quantum
probabilities as empirical frequencies
\cite{Zurek:1998:re,Deutsch:1999:tz}. However, as it has been pointed
out before \cite{Zeh:1996:gy,Zurek:2002:ii,Schlosshauer:2003:tv}, this
argument cannot yield a noncircular derivation of the Born rule, since
the formalism (in particular, the trace operation) and interpretation
of reduced density matrices presume this rule.

The solution to the problem of understanding the meaning of
probabilities and of deriving Born's rule in a relative-state
framework must therefore be sought on a much more fundamental level of
quantum mechanics. Since this framework presumes nothing besides the
unitarily evolving state vector itself, the solution should preferably
be derived solely from properties of this quantum state. However,
while we would like to assign probabilities to ``outcomes of
measurements'' on a local system (i.e., probabilities for the system
to be found in a certain state), the global quantum state usually
contains a high degree of environmental entanglement, i.e., there
exists no state vector that could be assigned to the local system
alone.  Still, we obviously talk regularly of the ``state of the
system,'' and we must therefore distinguish this notion of state from
the quantum state vector itself. Following the relative-state
viewpoint, the local ``events'' of the system (or its possible
``states of the system'') are then typically identified with the
relative-state components of the global state vector in the Hilbert
subspace corresponding to the system.

The recent enormous advances in the field of quantum information
theory, especially in the understanding of the properties and
implications of quantum entanglement, have shed some light on how one
might proceed. Quantum information theory has established the notion
that quantum theory can be viewed as a description of what, and how
much, ``information'' Nature is willing to proliferate. For example, a
peculiar feature of quantum mechanics is that complete knowledge of
the global pure bipartite quantum state $\ket{\Psi} = \left(
  \ket{\alpha_1}\ket{\beta_1} + \ket{\alpha_2}\ket{\beta_2}\right)
/\sqrt{2}$ itself does not appear to contain information about the
``absolute'' state of one of the subsystems.  This hints at ways how a
concept of ``ignorance,'' and therefore of probability, may emerge
directly from the quantum feature of entanglement without any
classical counterpart.

This idea has recently been developed in a series of papers by Zurek
\cite{Zurek:2002:ii,Zurek:2003:rv,Zurek:2003:pl,Zurek:2004:yb},
leading to a proposal for a derivation of Born's rule. As pointed out
by the present author \cite{Schlosshauer:2003:ms,Schlosshauer:2003:tv}
and made more explicit in the most recent of Zurek's articles on this
topic \cite{Zurek:2004:yb}, the derivation is still based on certain
assumptions that are not contained in the basic measurement-free
relative-state framework of quantum mechanics. One might argue how
strong these assumptions are. Zurek himself, for example, considers
some of them to be ``facts'' and regards others as ``natural'' and
``modest'' \cite{Zurek:2004:yb}; a somewhat more critical position
with respect to some of the assumptions has been assumed by the
present \cite{Schlosshauer:2003:ms} and other authors
\cite{Barnum:2003:yb,Mohrhoff:2004:tv}.  Granted these
assumptions, however, we consider Zurek's proposal a very promising
approach towards a deeper understanding of the origin of quantum
probabilities, and we shall therefore outline the basic ideas and
assumptions in the following (a more detailed description and
discussion of the approach can be found in
\cite{Zurek:2002:ii,Schlosshauer:2003:ms,Schlosshauer:2003:tv,Zurek:2004:yb}).

Zurek's derivation is based on a particular symmetry property
(referred to as ``environment-assisted envariance,'' or ``envariance''
for short) of composite quantum states, which is used to infer
complete ignorance about the state of the subsystem. The derivation
relies on a study of the properties of a composite entangled state and
therefore intrinsically requires the decomposition of the Hilbert
space into subsystems and the usual tensor-product structure. The core
result to be established is the following. Given a bipartite product
Hilbert space $H_{\mathcal{S}_1} \otimes H_{\mathcal{S}_2}$ and a completely
known composite pure state in the diagonal Schmidt decomposition
\be \label{eq:comp}
\ket{\Psi} = \left( e^{i\varphi_1} \ket{\alpha_1}_1 \ket{\beta_1}_2
+ e^{i\varphi_2} \ket{\alpha_2}_1 \ket{\beta_2}_2 \right) / \sqrt{2},
\ee
where the $\ket{\alpha_i}_1$ and $\ket{\beta_i}_2$ are orthonormal
basis vectors that span the Hilbert spaces $H_{\mathcal{S}_1}$ and
$H_{\mathcal{S}_2}$, the probabilities of obtaining either one of the
relative states $\ket{\alpha_1}_1$ and $\ket{\alpha_2}_1$ (identified
with the ``events'' of interest to which probabilities are to be
assigned \cite[p.~12]{Zurek:2003:pl}; see also
\cite{Schlosshauer:2003:ms}) are equal. Given this result,
generalizations to higher-dimensional Hilbert spaces and to the case
of unequal absolute values of the Schmidt coefficients in
Eq.~\eqref{eq:comp} can be achieved in a rather straightforward way
\cite{Zurek:2004:yb}.

This result is established in two key steps. First, a few simple
assumptions (Zurek's ``facts'' \cite{Zurek:2004:yb}) are introduced
that connect the global quantum state $\ket{\Psi}$,
Eq.~\eqref{eq:comp}, to the ``state of the system'' $\mathcal{S}_1$.
This is necessary because, as mentioned above, the global quantum
state is all that the pure state-vector formalism of quantum mechanics
provides for the description of a bipartite system containing
entanglement.  The following assumptions are made about the ``state of
the system'' $\mathcal{S}_1$. First, this state is completely
determined by the global quantum state, Eq.~\eqref{eq:comp}; second,
it specifies all measurable properties of $\mathcal{S}_1$, including
probabilities of outcomes of measurements on $\mathcal{S}_1$; and
third, unitary transformations can change it only if they act on
$\mathcal{S}_1$ (see \cite{Schlosshauer:2003:ms} for a discussion
of this last assumption).

Granted these three assumptions, one can show that measurable
properties of $\mathcal{S}_1$ can depend neither 

\bn

\item on the phases $\varphi_i$ in Eq.~\eqref{eq:comp}, such that we can
  assume the simplified form
\be
\ket{\Psi} = \left( \ket{\alpha_1}_1 \ket{\beta_1}_2
  + \ket{\alpha_2}_1 \ket{\beta_2}_2 \right) / \sqrt{2}
\ee
for our purpose of discussing probabilities associated with
$\mathcal{S}_1$;

\item nor on whether $\ket{\alpha_1}_1$ is paired with
  $\ket{\beta_1}_2$ or $\ket{\beta_2}_2$, i.e., the unitary
  transformation acting on $\mathcal{S}_1$ that changes the quantum
  state vector
\be
\ket{\Psi} = \left( \ket{\alpha_1}_1 \ket{\beta_1}_2
  + \ket{\alpha_2}_1 \ket{\beta_2}_2 \right) / \sqrt{2}
\ee
into
\be
\ket{\Psi'} = \left( \ket{\alpha_2}_1 \ket{\beta_1}_2
  + \ket{\alpha_1}_1 \ket{\beta_2}_2 \right) / \sqrt{2}
\ee
cannot have altered the state of $\mathcal{S}_1$.

\en

In a way, result (2) already indicates a feature of ignorance about
the state of $\mathcal{S}_1$, since interchanging the potential
``outcomes'' $\ket{\alpha_i}_1$ through local operations performed on
$\mathcal{S}_1$ does not change any measurable properties of
$\mathcal{S}_1$ and can therefore be viewed as leading to a form of
``objective indifference'' among these outcomes. It is important to
note that this effect is crucially dependent on the feature of
entanglement. In a nonentangled pure state of the form $\ket{\Phi} =
\left( \ket{\phi_1} + e^{i\varphi} \ket{\phi_2} \right) \sqrt{2}$, the
phase $\varphi$ must of course not be ignored (and would be measurable
in a suitable interference experiment), and therefore the system
described by the ``swapped'' state $\ket{\Phi'} = \left( \ket{\phi_2}
  + e^{i\varphi} \ket{\phi_1} \right) \sqrt{2}$ is clearly physically
different from that represented by the original state $\ket{\Phi}$.

To make the above argument more precise, in the second key step of the
derivation, the notion of probabilities of the outcomes
$\ket{\alpha_i}_1$ in a measurement performed on $\mathcal{S}_1$
(previously only subsummed under the general heading ``measurable
properties of $\mathcal{S}_1$'') is now explicitly connected to the
global state vector via an additional assumption. In
\cite{Zurek:2004:yb}, Zurek offers three possible choices for
this assumption, of which we should quote one (see also
\cite{Barnum:2003:yb}). Namely, it is assumed that the form of
the Schmidt product states $\ket{\alpha_i}_1\ket{\beta_i}_2$ appearing
in Eq.~\eqref{eq:comp} implies that the probabilities for
$\ket{\alpha_i}_1$ and $\ket{\beta_i}_2$ must be equal. Given this
assumption and using result (2) above, it can be readily established
\cite{Schlosshauer:2003:ms,Schlosshauer:2003:tv,Zurek:2004:yb} that
the probabilities for $\ket{\alpha_1}_1$ and $\ket{\alpha_2}_1$ must
be equal, thus completing the derivation.

As we have pointed out elsewhere \cite{Schlosshauer:2003:ms}, the need
for the final assumption may be considered a reflection of the
well-worn phrase that a transition from a nonprobabilistic theory
(such as quantum mechanics solely based on deterministically evolving
state vectors) to a probabilistic theory (that refers to
``probabilities of outcomes of local measurements'') requires, at some
stage, to ``put probabilities in to get probabilities out.'' However,
in the quantum setting, this introduction of a probability concept has
a far more objective character than in the classical case. While in
the latter setting probabilities refer to subjective ignorance in
spite of the existence of a well-defined state (see also
Sec.~\ref{sec:object}), in the quantum case all that is available,
namely, the global entangled quantum state, is perfectly known. The
objectivity of ignorance in quantum mechanics can thus be viewed as a
consequence of a form of ``complementarity'' between local and global
observables \cite{Zurek:2004:yb} and could help explain the
fundamental need for a probabilistic description in the quantum
setting despite the deterministic evolution of the global state
vector.

It is the great merit of Zurek's proposal to have emphasized this
objective character of quantum probabilities arising from the feature
of quantum entanglement. While the precise role and importance of the
assumptions entering the derivation as well as the generality of the
approach (given, e.g., the focus on Schmidt decompositions) would
benefit from further discussion and analysis, the approach definitely
sheds an interesting and new light on the nature of quantum
probabilities.

\section{Objectification of observables in a relative-state framework}
\label{sec:object}

A characteristic feature of classical physics is the fact that the
state of a system can be found out and agreed upon by many independent
observers (with all of them initially completely ignorant about the
state) without disturbing this state. In this sense, classical states
preexist objectively, resulting in our notion of ``classical
reality.'' In contrast, as is well known, measurements on a closed
quantum system will in general alter its state---unless, of course,
the observer chooses to measure, by pure luck or prior knowledge, an
observable with an eigenstate that coincides with the state of the
system.  It is therefore impossible to regard quantum states of a
closed system as existing in the way that classical states do.  This raises
the question of how classical reality emerges from within the quantum
substrate, i.e., how observables are ``objectified'' in the above
sense.

In a first step, the decoherence program, in particular the stability
criterion and the more general formalism of the ``predictability
sieve''
\cite{Zurek:1981:dd,Zurek:1982:tv,Zurek:1993:pu,Zurek:1998:re,Zurek:2002:ii,%
  Schlosshauer:2003:tv} (see also Sec.~\ref{sec:squid-prefbasis}), has
provided an answer to the question of why only a certain subset of the
possible states in the Hilbert space of the system are actually
observed. Taking into account the openness of the system and the form
of the system-environment interaction is crucial in determining a set
of preferred stable states of the system. This supplies an elegant and
physically motivated solution to the problem of the preferred basis,
an issue that has often been used to challenge the feasibility of
relative-state interpretations \cite{Kent:1990:nm,Stapp:2002:pc}.
Nonetheless, the problem sketched in the previous paragraph remains,
as any direct measurement performed on the system would, in general,
still alter the state of the system.

The important next step is therefore to realize that in most (if not
all) cases observers gather information about the state of a system
through indirect observations, namely, by intercepting fragments of
environmental degrees of freedom that have interacted with the system
in the past and thus carry information about the state of the system
\cite{Zurek:1993:pu,Zurek:1998:re,Zurek:2000:tr,Zurek:2002:ii}.
Probably the most common example for such an indirect acquisition of
information is the visual registration of photons that have scattered
off from the object of interest (see also Sec.~\ref{sec:chain}).
Similiar to the case of decoherence, the recognition of the openness
of quantum systems is therefore crucial.  However, the role of the
environment is now broadened, namely, from the selection of preferred
states for the system of interest and the dislocalization of local
phase coherence, to the transmission of information about the state of
the system. The idea is then to show how, and which, information is both
redundantly and robustly stored in a large number of distinct
fragments of the environment in such a way that multiple observers can
retrieve this information without disturbing the state of the system,
thereby achieving effective classicality of the state. 

This approach has recently been developed under the labels of
``environment as a witness'' (i.e., the recognition of the role of the
environment as a communication channel) and ``quantum Darwinism''
(namely, the study of what information about the system can be stably
stored and proliferated by the enviroment)
\cite{Zurek:1993:pu,Zurek:1998:re,Zurek:2002:ii,Zurek:2003:pl,Ollivier:2003:za,%
  Ollivier:2004:im,Blume:2004:oo,Blume:2005:oo}.  To explicitly
quantify the degree of completeness and redundancy of information
imprinted on the enviroment, the measure of (classical
\cite{Ollivier:2003:za,Ollivier:2004:im} or quantum
\cite{Zurek:2002:ii,Blume:2004:oo,Blume:2005:oo}) mutual information
has usually been used.  Roughly speaking, this quantity represents the
amount of information (expressed in terms of Shannon or von Neumann
entropies) about the system $\mathcal{S}$ that can be acquired by
measuring (a fragment of) the environment $\mathcal{E}$.  Note that
the amount of information contained in each fragment is always
somewhat less \cite{Blume:2005:oo} than the maximum information
provided by the system itself (as given by the von Neumann entropy of
the system).

The measure of classical mutual information is based on the choice of
particular observables of $\mathcal{S}$ and $\mathcal{E}$ and
quantifies how well one can predict the outcome of a measurement of a
given observable of $\mathcal{S}$ by measuring some observable on a
fraction of $\mathcal{E}$ \cite{Ollivier:2003:za,Ollivier:2004:im}.
The quantum mutual information
$\mathcal{I}_{\mathcal{S}:\mathcal{E}}$, used in more recent studies
\cite{Zurek:2002:ii,Blume:2004:oo,Blume:2005:oo}, can be viewed as a
generalization of classical mutual information and is defined as
$\mathcal{I}_{\mathcal{S}:\mathcal{E}} = H(\mathcal{S}) +
H(\mathcal{E}) - H(\mathcal{SE})$, where $H(\rho) = -\text{Tr} (\rho
\log \rho)$ is the von Neumann entropy. Thus
$\mathcal{I}_{\mathcal{S}:\mathcal{E}}$ measures the amount of entropy
produced by destroying all correlations between $\mathcal{S}$ and
$\mathcal{E}$, i.e., it quantifies the degree of correlations between
$\mathcal{S}$ and $\mathcal{E}$. Results derived from these measures
have thus far been found to be sufficiently robust with respect to the
particular
choice of measure \cite{Ollivier:2003:za,Ollivier:2004:im,Zurek:2002:ii,%
  Blume:2004:oo,Blume:2005:oo}, although a more detailed analysis of
this issue is underway \cite{Blume:2005:oo}.

It has been found that the observable of the system that can be
imprinted most completely and redundantly in many distinct subsets of
the environment coincides with the ``pointer'' observable selected by
the system-environment interaction (i.e., by the stability criterion
of decoherence)
\cite{Ollivier:2003:za,Ollivier:2004:im,Blume:2004:oo,Blume:2005:oo}.
Conversely, most other states do not seem to be redundantly storable.
Thus it suffices to probe a comparably very small fraction of the
environment to infer a large amount of the maximum information about
the pointer state of the system.  On the other hand, if the observer
tried to measure other observables on the same fragment, he would
learn virtually nothing, as information about the corresponding
observables of the system is not redundantly stored.  Thus the
``pointer'' states of the system play a twofold role: They are the
states least perturbed by the interaction with the environment, and
they are the states that can be most easily found out, without
disturbing the system, by probing environmental degrees of freedom.
Since the same information about the pointer observable is stored
independently in many fragments of the environment, multiple observers
can measure this observable on different fragments and will
automatically agree on the findings. In this sense, one can ascribe
(effective) objective existence to the pointer states.

The research into the objectification of observables along the lines
outlined in this section is only in its beginnings. Important aspects,
such as the explicit dynamical evolution of the objectification
process \cite{Ollivier:2004:im} and the role of the assumptions and
definitions in the current treatments of the ``objectification through
redundancy'' idea, are currently still under investigation, as are
studies involving more detailed and realistic system-environment
models.  However, it should have become clear that the approach of
departing from the closed-system view and of describing observations
as the interception of information that is redundantly and robustly
stored in the environment, represents a very promising candidate for a
purely quantum-mechanical account of the emergence of classical
reality from the quantum domain.

\section{Decoherence in the perceptive and cognitive apparatus} \label{sec:brain}

If, motivated by the results of the experiments described in
Sec.~\ref{sec:experiments}, we assume the universal validity of the
Schr\"odinger equation, we immediately face two related consequences:

\bn

\item We ought to reconcile this assumption with our perception
  of definite states in the macroworld, since now there is no
  underlying stochastic mechanism (of whatever nature) that would
  select, in an objective manner, a particular ``outcome'' among the
  terms in a superposition of, say, spatially localized wave packets.
  There exists not only a multitude, but also interference effects
  between them.
  
\item If Schr\"odinger dynamics are universal, it is reasonable
  (at least from a scientifically reasonable functionalist's
  standpoint) to also describe observers with their perceptive and
  cognitive apparatuses---including even what could be grouped
  together under the rather vague term of ``consciousness''
  \cite{Neumann:1932:gq,Wigner:1962:iu,Stapp:1993:mm,Zeh:2000:rr}---by
  unitarily evolving wave functions.

\en

Both consequences follow quite naturally from the assumption of
universally exact [consequence (1)] and universally applicable
[consequence (2)] Schr\"odinger dynamics.  Quite generally, the
preferred strategy would be to treat them jointly: Solving the
``measurement problem,'' that is, consequence (1), posed by the
assumption of a purely unitary quantum theory, by applying this very
theory to the observer, i.e., consequence (2). If successful, this
would lead to a ``subjective'' resolution of the measurement problem,
i.e., to a quantum-mechanical account of why we, as observers,
perceive {\it definite} states in {\it specific} bases, rather than
superpositions of these states.  In the opinion of this author
\cite{Schlosshauer:2003:tv} and of others (see, e.g.,
\cite{Zeh:1973:wq,Zurek:1998:re,Espagnat:2000:uy,Zeh:2000:rr,Zurek:2002:ii}),
this would also represent a sufficient solution to the problem.

\subsection{General remarks}

First of all, on a rather philosophical sidenote, it is clear that the
familiar concepts of the world of our experience are expressed in
terms of the observed specific definite states.  We do not even {\it
  have} a concept available for what a state describing a
superposition of an alive and dead cat would represent, because we
have never observed such a state. While such a Schr\"odinger cat might
seem exotic, we have seen that quite analogous states are realizable
in the laboratory --- for example, in terms of superpositions of
currents running in opposite directions in SQUIDs.  As we have argued
in Sec.~\ref{sec:interpret-superpos}, the only way we can access such
superpositions in terms of our concepts (and not just in mathematical
terms) is through the definite current states $\ket{L}$ and $\ket{R}$
that are observable as individual preferred states of the system upon
measurement.

Furthermore, it is virtually indisputable that we must describe all
observations in terms of physical interactions between the observed
system and the observer, i.e., by means of an appropriate interaction
Hamiltonian $\widehat{H}_\text{int}$.  Such interactions do not have
to be, and usually are not, direct. For example, the probably most
common type of observation involves the interception of a number of
photons that have interacted with the object of interest in the past
and whose state is thus entangled with the state of the object.  These
photons then contain indirect and redundantly coded information about
the object that can be revealed without significantly disturbing the
state of the object (see Sec.~\ref{sec:object}).

If the perception of definiteness is not introduced as an extraneous
postulate, but is rather understood as emerging from the unitary
quantum formalism itself when observations and observers are described
in physical terms, it is inevitable that attempts have to be made to
analyze the cognitive apparatus in quantum-mechanical terms.  It is
clear that giving such an account of subjective definiteness by
referring to the physical structure of observers cannot share the
mathematical compactness and exactness of axiomatically introduced
rules that enforce definiteness on a fundamental level of the theory.
However, it is important to note that, given the paramount role of
observations in quantum mechanics (mostly owing to the fact that, in
general, states do not pre-exist in a classical sense), postulating
such ``exact'' rules is tantamount to simply avoiding a physical
analysis of crucial and objective (that is, interpretation-neutral)
physical processes (cf.\ Kent's objections to ``many-worlds''
interpretations \cite{Kent:1990:nm} and Wallace's defense
\cite{Wallace:2003:iz}).

If a purely unitary time evolution is assumed and observations are
modeled as physical interactions, the conclusion of the existence of
quantum-mechanical superpositions of brain states corresponding to the
different ``outcomes'' of observations is inescapable. Individual
perceptions are represented by certain neuronal resting/firing
patterns in the brain (see \cite{Donald:1995:lk,Donald:2002:um}
for more precise definitions of this relationship).  As we shall
discuss in the next section, superpositions of resting and firing
states of a neuron are extremely sensitive to environmental
decoherence, with the resting and firing states forming the robust
neuronal states. These states can thus be identified with ``record
states'' that are capable of robustly encoding information in spite of
environmental interactions \cite{Zurek:1998:re,Zurek:2004:yb}. As a
consequence of the practically irreversible dislocalization of phase
relations between these record states through entanglement with the
environment, a dynamical decoupling of these states results.  This
process represents an objective branching process due to physical
interactions between subsystems and with the environment.

The remaining question is then how to relate this objective branching
to the perceived subjective ``branches of consciousness,'' i.e.,
collective memory states, or ``minds'' (von Neumann's principle of the
``psycho-physical parallelism'' \cite{Neumann:1932:gq}). Of course,
the existence (and therefore the locality) of consciousness cannot
actually be {\em derived} from the quantum-mechanical formalism. This
has led some authors to conclude that the question of the relationship
between subjective experience and its physical correlates can only be
fully answered through the introduction of new physical laws
\cite{Donald:2002:um}.  However, in the opinion of this and other
authors (see, for example, \cite{Zeh:2000:rr,Zeh:2004:zm}), it is an
entirely viable (if not compelling) strategy to postulate, within the formalism, the
existence of consciousness based on the empirical fact of decohering
wavefunction components in neuronal processes, by associating the
robust components of the global wave function labelled by the
decohered neuronal states with dynamically autonomous observers
\cite{Zeh:1970:yt,Zeh:1973:wq,Lockwood:1996:pu,Zurek:1998:re,%
Zeh:2000:rr,Zurek:2002:ii,Zeh:2004:zm,Zurek:2004:yb}.

Due to the absence of more concrete theoretical and experimental
insight into the physical underpinnings of the cognitive apparatus
with its associated complex entities such as the ``mind,''
``consciousness,'' and even the comparably basic ``record states,''
the above brief account of how subjective definiteness may emerge from
purely unitary quantum mechanics must (at least for now) remain
inherently somewhat vague and nontechnical. Fortunately, however, the
main points of the argument are quite independent of, say, the precise
details of the structures and dynamics of the information-processing
cognitive entities, since the ubiquity and effectiveness of
decoherence is likely to lead to very robust results.  We shall
therefore turn, in the next section, to concrete estimates for 
decoherence rates in neurons.

\subsection{Decoherence of neuronal superpositions}

The extremely complex network of about $10^{11}$ interacting neurons
in the brain undoubtly comprises a major part of the cognitive
machinery used for processing and storing of information obtained from
sensory input.  Computer models of such neuronal networks (employing a
massively parallel interconnected web of ``switches'' that are turned
on and off depending on some, typically nonlinear, activation
function) can exhibit rich and complex behavior similiar to that
encountered in cognitive processes.\footnote{However, as Donald
  \cite{Donald:2002:um} has pointed out, the brain should not be
  thought of as a deterministic classical computer with a predictable
  input/output pattern, since synaptic transmissions have a fairly
  high failure rate due to the complexity of the underlying biological
  processes.  The large number of about $10^{14}$ synapses in the
  human brain, with each neuron firing in average several times per
  second, inevitably leads to a high degree of unpredictability on the
  ``everyday level'' that is much more significant than effects due to
  pure quantum uncertainties.  } In particular, it is reasonable to
identify the ``record states'' mentioned above with individual neurons
or neuronal clusters.  One might conjecture that ultimately all
cognitive processes (and thus presumably also our perception of
consciousness) are due to neuronal activity.

\begin{figure}
\begin{center}
\includegraphics[scale=.3]{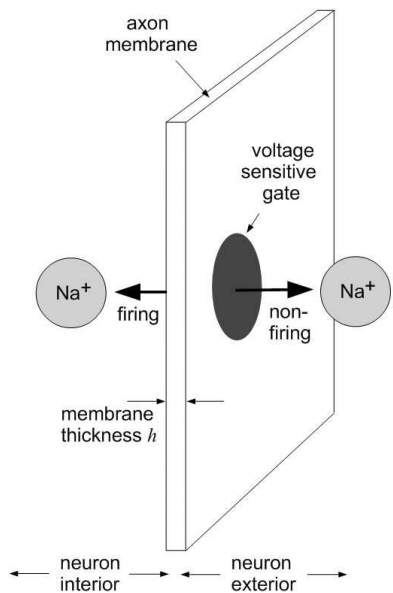}
\end{center}
\caption[Schematic illustration of the axon
membrane of a neuron]{\label{fig:neuron} Schematic illustration of the
  axon membrane of a neuron. The firing of the neuron corresponds to a
  net flow of $N \sim 10^6$ sodium ions into the inside of the axon.
  Superpositions of firing and non-firing neuronal states (i.e., of
  $N$ ions being in a spatial superposition of inside and outside the
  membrane) are decohered on a time scale of about $10^{-20}$~s
  \cite{Tegmark:2000:wz}. }
\end{figure}

Thus the importance of a quantitative investigation of decoherence in
neuronal states should be clear. Tegmark \cite{Tegmark:2000:wz} has
estimated decoherence rates for a superposition of a firing and
non-firing neuron in the brain. The firing is represented by a large
number $N \sim 10^6$ \cite{Tegmark:2000:wz} of Na$^+$ ions moving
across the membrane into the inside of the axon
(Fig.~\ref{fig:neuron}). Thus, a superposition of a firing and
nonfiring neuron corresponds to a spatial superposition involving
$O(N)$ Na$^+$ ions.

The extensive difference $\mathcal{S}_\text{ext}$ (see
Sec.~\ref{sec:measures}) can then be estimated to be on the order of
$10^2$--$10^3$, given by a small multiple of the thickness $h \sim
10$~nm of the axon membrane separating the inside and outside regions,
relative to the size of a Na$^+$ ion, which is on the order of 0.1~nm.
While this value for $\mathcal{S}_\text{ext}$ is comparably small, the
degree of entanglement $\mathcal{S}_\text{ent}$ is somewhat closer to
the values listed in Table~\ref{tab:degrees}.  Taking it to be equal
to the number of microsopic constituents, we obtain
$\mathcal{S}_\text{ent} \sim 3 \times 10^7$. Thus a neuron being in a
superposition of firing and resting quite clearly falls into the
macrosopic category.

The decoherence rates for this superposition as estimated by Tegmark
are, as expected, extremely fast. The three main sources of
decoherence in this case, namely, ion-ion scattering, ion-water
collisions, and long-range Coulomb interactions due to nearby ions,
all result in decoherence times on the order of $10^{-20}$~s. 

One obvious implication of fast neuronal decoherence is that coherent
superpositions in neurons could never be sustained long enough to
allow for some form of quantum computation. This result appears to be
much more clearly established than an answer to the question of
whether the relevant decoherence times are long enough to allow for
quantum computation in microtubules (dynamically active structures
that are a dominant part of the cytoskeleton, i.e., the internal
scaffolding of cells). Suggestions in the positive, including the
association of such quantum computations with the emergence of
consciousness, have been put forward in
\cite{Penrose:1994:mm,Hameroff:1996:im,Hameroff:1996:iy},
criticized in \cite{Tegmark:2000:wz}, subsequently defended in
\cite{Hagan:2002:th}, and further evaluated in
\cite{Rosa:2004:um} (see also \cite{Stapp:2000:yy}).

However, the question much more relevant to the theme of this paper
concerns the implications of neuronal decoherence for a
decoherence-based account of subjective definiteness in unitary
quantum mechanics --- i.e., for a subjective resolution of the
``measurement problem.'' To this extent,  let us in the following
discuss a simple step-by-step quantum-mechanical account of the chain
of interactions leading to the recording of a visual event in the brain.

\subsection{Schematic sketch of the chain of interactions in visual perception
  and cognition} \label{sec:chain}

Suppose that a small number of photons interact with an object
$\mathcal{O}$ described by a pure-state superposition of two
macrosopically distinct positions. This step already can be viewed as
an environmental decoherence process, where now, however, the
environment assumes a crucial role as a carrier of information (see
Sec.~\ref{sec:object}). Due to entanglement, the combined
object-photon system will be described by a superposition of the form
\be
\ket{\Psi_\mathcal{OP}} = \frac{1}{\sqrt{2}} \bigl(
\ket{\omega_1}_\mathcal{O} \ket{\phi_1}_\mathcal{P} +
\ket{\omega_2}_\mathcal{O} \ket{\phi_2}_\mathcal{P} \bigr),
\ee
where $\omega_i$, $i=1,2$, are the two distinct (small) spatial
regions associated with the object, and $\ket{\phi_i}_\mathcal{P}$
denote the corresponding classically distinct collective photon
states. A conceptually similiar arrangement on the mesosopic scale has
explicitly been studied in experiments involving a single rubidium
atom (representing the object) in a superposition of two internal
levels and entangled with a cavity radiation mode (corresponding to
the collection of photons) \cite{Brune:1996:om,Raimond:1997:um}.

Initial detection of such a collection of photons in the (human) eye
is associated with rhodopsin molecules in the retina.  Due to their
mesoscopic properties, rhodopsin molecules are subject to strong
decoherence, such that already at this stage the influence of the
environment will have preselected the robust states
$\ket{\rho_i}_\mathcal{R}$ of the rhodopsin molecule, which correspond
to certain photon detection events $\ket{\phi_i}_\mathcal{P}$.  The
total state $\ket{\Psi_\mathcal{OPR}}$ will then be given by
\be
\ket{\Psi_\mathcal{OPR}} = \frac{1}{\sqrt{2}} \bigl(
\ket{\omega_1}_\mathcal{O} \ket{\phi_1}_\mathcal{P}
\ket{\rho_1}_\mathcal{R} +
\ket{\omega_2}_\mathcal{O} \ket{\phi_2}_\mathcal{P}
\ket{\rho_2}_\mathcal{R} \bigr),
\ee
i.e., the photon-rhodopsin interaction should lead to an (albeit, due
to the influence of decoherence, very fragile) superposition of the
different biochemically distinct states $\ket{\rho_i}_\mathcal{R}$ of
the rhodopsin molecule.\footnote{A search for experimental evidence
  for such superpositions has been suggested in
  \cite{Shimony:1998:yy}; for an experimental proposal, see
  \cite{Hilaire:2002:ya}. Cf.\ also \cite{Thaheld:2005:om}
  for an (unconvincing) suggestion that the visual apparatus itself
  might trigger a physical collapse.} These relative states can then be
expected to be further correlated with the appropriate states
$\ket{\nu_i}_\mathcal{N}$ of neuronal arrays that are mainly located
in the primary visual area in the occipital lobe of the brain. Suppose
that the two ``events'' represented by the two distinct states
$\ket{\rho_i}_\mathcal{R}$ of the rhodopsin molecule (corresponding to
the different photon states $\ket{\phi_i}_\mathcal{P}$ that in turn
carry information about the two distinct spatial regions $\omega_i$ of
the object) are encoded by the states $\ket{\nu_i}_\mathcal{N}$,
$i=1,2$, describing the same collection of $N$ neurons in two
different firing/resting patterns.

As a simple example, let us take $N=3$, and $\ket{\nu_1}_\mathcal{N} =
\ket{1}_{\mathcal{N}_1} \ket{0}_{\mathcal{N}_2}
\ket{1}_{\mathcal{N}_3}$ and $\ket{\nu_2}_\mathcal{N} =
\ket{0}_{\mathcal{N}_1} \ket{1}_{\mathcal{N}_2}
\ket{1}_{\mathcal{N}_3}$, where $\ket{0}_{\mathcal{N}_i}$ and
$\ket{0}_{\mathcal{N}_i}$ denote, respectively, the resting and firing
state of the $i$th neuron. Then the combined state
$\ket{\Psi_\mathcal{OPRN}}$ will be given by
\begin{multline}
\ket{\Psi_\mathcal{OPRN}} = \frac{1}{\sqrt{2}} \bigl(
\ket{\omega_1}_\mathcal{O} \ket{\phi_1}_\mathcal{P}
\ket{\rho_1}_\mathcal{R} \ket{1}_{\mathcal{N}_1} \ket{0}_{\mathcal{N}_2}
\ket{1}_{\mathcal{N}_3} 
\\ + \ket{\omega_2}_\mathcal{O} \ket{\phi_2}_\mathcal{P}
\ket{\rho_2}_\mathcal{R}  \ket{0}_{\mathcal{N}_1} \ket{1}_{\mathcal{N}_2}
\ket{1}_{\mathcal{N}_3} \bigr).
\end{multline}
The extreme fast decoherence rate for the neurons 1 and 2 being in a
superposition of firing and resting will lead to a practically
irreversible dynamical decoupling of the two branches that now
describe two distinct ``outcomes'' encoded by
$\ket{\nu_i}_\mathcal{N}$. We may then identify these states with the
basic memory states, although, strictly speaking, the physical process
of actual information storage in the brain (i.e., learning) occurs
only in two subsequent stages \cite{Kandel:2000:tr}. First, in form of
short-term memory, believed to be due to certain biochemical and
electrical interactions between neurons. Second, as long-term memory
that is based on actual structural changes in the brain
(``neuroplasticity''), most notably, due to the formation of new
connections (synapses) between neurons and due to internal changes in
the synaptic regions in individual neurons.

However, since all these processes will again be subject to strong
decoherence, the essence of our argument is not altered: The states in
a superposition of neuronal firing patterns will rapidly entangle with
approximately orthogonal (i.e., macrosopically distinguishable) states
of the environment and thus lead to the formation of locally
noninterfering (that is, dynamically autonomous) branches labelled by
these ``outcome'' states. Regardless of the precise physical,
chemical, biological, psychological, etc., details of perceptive and
cognitive activity, it is quite clear that decoherence effects are
likely to be sufficient to explain the emergence of a subjective
perception of single outcomes, represented by stable, ``classical,''
record states, from a (by all accounts macrosopic) global
superposition.

\section{Discussion and outlook} \label{sec:discussion}

We have analyzed three important experimental domains---namely,
SQUIDs, molecular diffraction, and Bose-Einstein condensation---that
have demonstrated (or at least have come very close to demonstrating)
the existence of superpositions of states that can be considered
macrosopically distinct in comparison with the microsopic states
``typically'' treated in quantum mechanics. These experiments have
provided powerful examples for the validity of unitary Schr\"odinger
dynamics and the superposition principle on increasingly large length
scales. They have also shown how the fragility of macrosopic
superpositions can be precisely understood and controlled in terms of
environmental interactions and the resulting decoherence effects.

Of course, these experiments do not {\em falsify} the possibility that
the Schr\"odinger equation might not be exact under all circumstances.
In fact, no finite number of experiments that show the validity of
unitary dynamics could ever do. To do so, a ``positive-test''
experiment would be needed that could explicitly demonstrate nonlinear
deviations from the Schr\"odinger equation.  Leggett
\cite{Leggett:1980:yt,Leggett:2002:uy} has presented a Bell-type
inequality that would be obeyed by what he calls the class of
``macrorealistic theories,'' while it would be violated by the
predictions of purely unitary quantum mechanics. The
``macrorealistic'' class is defined to represent all theories in which
macrosopic systems are always in a single definite state among a
collection of possible macrosopically distinct states, and in which
this definite state can be found out without perturbing the state and
dynamics of the system.  So one might, at least in principle, through
suitable experiments be able to exclude either any such macrorealistic
theory or the universal validity of the Schr\"odinger equation. Such a
strategy would be similiar in spirit to the tests of Bell's
inequalities, which rule out a large class of, if not all, local
realistic theories.  (See Sec.~6 of \cite{Leggett:2002:uy} and
references therein for some first ideas in this direction.)

At the current stage, however, it is the opinion of the present author
that, in  absence of any positive evidence for deviations from
unitary dynamics, combined with the continued experimental
verification of increasingly large ``Schr\"odinger cats'' (whose time
evolution, including decoherence effects, is in perfect agreement with
unitary dynamics), it appears to be not only reasonable, but moreover
compelling, to entertain the possibility of a universally exact
Schr\"odinger equation seriously and to fully explore the consequences
of this assumption.

The experiments described in this paper have demonstrated rapid
progress in achieving, controlling, and observing superpositions of
increasingly distinct states. Experiments involving superpositions of
classically distinguishable states of a few photons
\cite{Brune:1996:om,Raimond:1997:um} have been followed by collective
superpositions of $10^9$ electron pairs in SQUIDs and double
slit--type experiments using massive molecules with a large number of
degrees of freedom. It is only a matter of time until
number-difference superpositions involving on the order of $10^7$
rubidium atoms will be experimentally realized in BECs. It is rather
unlikely that this progress towards experimental evidence for
increasingly large superpositions will encounter any fundamental
boundaries in the near future. As we have seen, the main limit seems
to be given by the ability to shield the system sufficiently from the
decohering influence of the environment. This limit is open to precise
quantitative analysis.

In view of this situation, we may now legitimately ask what the next
steps in solidifying the empirical support for a purely unitary
quantum theory and its consequences might ideally look like.  To this
extent, we remark that superpositions of macrosopically distinct
states that refer to biological (and, even more so, animate) objects
seem to have been considered as particular ``paradoxical'' --- after
all, Schr\"odinger chose a cat to illustrate his famous {\em Gedanken}
experiment.  This attitude may be traced back to several reasons. For
example:

\bn

\item The ``distinctness'' between the states referring to biological
  objects is usually extremely complex. Not only is the number of
  physical, chemical, biological, etc., differences between a dead and
  alive cat overwhelmingly vast, even two functionally different
  states of a simple biological molecule will be distinct in a large
  number of features.  By contrast, in the examples involving
  inanimate objects, such as BECs and SQUIDs, the states in the
  superposition usually differ only in a single physical
  quantity, such as  total angular momentum or magnetic moment.
  
\item While we might be willing to accept the existence of an
  ``exotic'' superposition under extreme physical conditions (such as
  superconducting currents in a bulk of matter cooled down to
  temperatures close to absolute zero), biological objects reside in
  the parameter regime characteristic of the world of our everyday
  experience.
  
\item If the superposition principle is applied to human observers
  (specifically, superpositions of ``states of consciousness,'' etc.),
  we feel that our most basic intuition about possessing a unique
  identity has been infringed upon.
  
\en
  
Especially in light of  the first two arguments, the molecular
diffraction experiments appear to be the most ``natural'' realization
of superpositions of macrosopically distinct states. In fact, as
pointed out in Sec.~\ref{sec:diffrac-dec}, interference effects have
already be experimentally demonstrated for a biological molecule
\cite{Hackermueller:2002:wb}, and larger biological structures are
likely to follow \cite{Arndt:2002:bo,Hackermuller:2003:uu} (see also
Fig.~\ref{fig:extrapol}).

However, another interesting direction could also be taken from here.
As suggested for example in
\cite{Shimony:1998:yy,Leggett:2002:uy}, one might try to look
for interference effects between (and thus superpositions of) {\em
  biologically} distinct states of the same biomolecule, rather than
for the spatial superpositions demonstrated in the current
molecular-diffraction setups. While such experiments would be
considerably more difficult to carry out due to the required near--{\em
  in vivo} environmental conditions (room temperatures, presence of a
surrounding medium such as an aqueous solution, etc.), which would lead to very
strong decoherence effects, there does not seem to exist a fundamental
obstacle that would prevent one in principle from the realization of
such superpositions in a cleverly designed setup.

Experiments that would find some basic biological structure in a
superpositions of distinct states corresponding to different
biological ``inputs'' might in turn indicate the presence of a
superposition of input signals originating from the inanimate outside
world (e.g., a superposition of photon states entangled with spatially
distinct states of a single object --- see Sec.~\ref{sec:chain}).
They could also provide direct empirical evidence for consequences of
purely unitary quantum mechanics in the regime of more complex
structures that are part of conscious (human) observers, and might
therefore also ease the discomfort spelled out in item (3) above.

Given that experiments \cite{Hackermueller:2002:wb} have demonstrated
a splitting of the localized state of a biomolecule into ``branches''
corresponding to distinct paths, it would also be 
worth discussing, as Zeh \cite{Zeh:2000:rr} puts it,

\begin{quote}
  the consequences of similar \emph{Gedanken} experiments with objects
  carrying some primitive form of ``core consciousness'' --- including
  an elementary awareness of their path through the slits.
\end{quote}

In such a situation, after passage through the slits, the state of the
object would be described by a superposition of spatially distinct
trajectories. However, due to its awareness of the path, it would thus
also be in a superposition of multiple (local) ``states of
consciousness.''  Environmental scattering would then lead to
entanglement with path-encoding variables (decoherence), which hence
would also destroy interference effects between the ``branches of
consciousness,'' and thus the different paths would be ``experienced''
separately. In the absence of decoherence, it would be possible to
coherently recombine the branches into a single localized wavepacket
identical to the state before the passage through the slits.  It then
follows from the standard quantum-mechanical formalism that the
associated object then cannot have retained any ``memory'' of the path
taken before the recombination.  For related ideas using the example
of neutron interferometry, see \cite{Vaidmain:1998:zp}.

As it is well known, Bohr has repeatedly insisted on the fundamental
role of classical concepts (see, for example,
\cite{Bohr:1923:um,Bohr:1948:um}). The experimental evidence for
superpositions of macrosopically distinct states on increasingly large
length scales counters such a dictum. Superpositions appear to be
novel and individually existing states, often without any classical
counterparts.  Only the physical interactions between systems then
determine a particular decomposition into classical states from the
view of each particular system. Thus classical concepts are to be
understood as locally emergent in a relative-state sense and should no
longer claim a fundamental role in the physical theory.

We have already widely acknowledged, based on experimental evidence,
the fundamental nonlocality of the quantum world, in spite of the
utterly nonclassical implications. We also have obtained direct
evidence for the validity of unitary dynamics and the superposition
principle in all experiments conducted so far, although this has
forced us to again accept extremely nonclassical situations as
physical reality. Why not let these experiences guide us to extend our
willingness to entirely give up classical prejudice and instead
explore the consequences of a strictly unitary quantum theory embedded
into a minimal interpretive framework?  After all, exploring the
implications of pure quantum features to the largest possible extent
can in turn give us back the familiar ``classical'' notions of the
world of our experience. As we have discussed in this paper,
consequences of highly nonlocal quantum entanglement lead to the local
disappearance of quantum interference effects, may explain the origin
of probabilities in quantum mechanics, and are likely capable of
accounting for the objectification of observables and therefore the
emergence of effective classical reality, thus supplying the missing
pieces of the basic Everett theory that have frequently been been used
to challenge the viability of a relative state--type ``minimal
interpretation.''

\begin{acknowledgements}
  
The author thanks A.\ Fine and H.\ D.\ Zeh for fruitful
discussions and helpful comments.

\end{acknowledgements}

\bibliographystyle{apsrev}

\end{document}